\documentclass[letterpaper]{article}
\usepackage[utf8]{inputenc}
\usepackage{fullpage}

\usepackage[all]{nowidow}

\usepackage[square,numbers]{natbib}
\usepackage{graphicx}
\usepackage{float}
\usepackage{soul}
\usepackage[usenames, dvipsnames]{color}
\usepackage{textcomp}
\usepackage{ccaption}
\usepackage{hyperref}

\usepackage{amsmath,amsthm,amssymb,latexsym} 
\usepackage{bm}
\usepackage{esvect} 

\usepackage[left,mathlines]{lineno}

\usepackage{setspace}

\usepackage[english, greek]{babel} 

\usepackage{upgreek}

\newcommand{\delt}{$\updelta$}

\newcommand{\gam}{$\upgamma$}

\newcommand{\proj}{\mathbb{P}}
\newcommand{\norm}[1]{\left\lVert#1\right\rVert}
\newcommand{\floor}[1]{\left\lfloor{#1}\right\rfloor}
\newcommand{\setX}{\mathcal{X}}
\newcommand{\setY}{\mathcal{Y}}
\newcommand{\setZ}{\mathcal{Z}}

\newcommand{\setB}{\mathcal{B}}

\newcommand{\setD}{\mathcal{D}}
\newcommand{\setT}{\mathcal{T}}
\newcommand{\setV}{\mathcal{V}}
\newcommand{\setPC}{\mathcal{PC}}
\newcommand{\setMC}{\mathcal{MC}}
\newcommand{\bigO}{\mathcal{O}}

\DeclareMathOperator{\spanset}{span}
\DeclareMathOperator{\covar}{cov}
\DeclareMathOperator{\asinh}{sinh^{-1}}

\newcommand{\hlcy}[1]{{\textbf{#1}}}

\renewcommand{\footnote}[1]{{\sethlcolor{magenta}\hl{#1}}}

\usepackage{environ}
\NewEnviron{newalign}{
\begin{linenomath*}
\begin{align*}
    \BODY
\end{align*}
\end{linenomath*}
}

\title{\texorpdfstring{\textbf{Measuring inter-cluster similarities with Alpha Shape TRIangulation in loCal Subspaces (ASTRICS) facilitates visualization and clustering of high-dimensional data.}}{Measuring inter-cluster similarities with Alpha Shape TRIangulation in loCal Subspaces (ASTRICS) facilitates visualization and clustering of high-dimensional data.}}

\author{Joshua M.~Scurll\textsuperscript{1,2}}
\date{\normalsize 
\textsuperscript{1} Department of Mathematics and Institute of Applied Mathematics, 1984 Mathematics Road, University of British Columbia, Vancouver, British Columbia V6T 1Z2 Canada.\\ 
\textsuperscript{2} Email: \href{mailto:jscurll.ubc+astrics@gmail.com}{\nolinkurl{jscurll.ubc+astrics@gmail.com}}} 

\begin{document}

\selectlanguage{english}

\maketitle





\noindent
{\large \textbf{Abstract}}

\smallskip

\noindent
Clustering and visualizing high-dimensional (HD) data are important tasks in a variety of fields. For example, in bioinformatics, they are crucial for analyses of single-cell data such as mass cytometry (CyTOF) data.
Some of the most effective algorithms for clustering HD data are based on representing the data by nodes in a graph, with edges connecting neighbouring nodes according to some measure of similarity or distance.
However, users of graph-based algorithms are typically faced with the critical but challenging task of choosing the value of an input parameter that sets the size of neighbourhoods in the graph, e.g.~the number of nearest neighbours to which to connect each node or a threshold distance for connecting nodes.
The burden on the user could be alleviated by a measure of inter-node similarity that can have value 0 for dissimilar nodes without requiring any user-defined parameters or thresholds. This would determine the neighbourhoods automatically while still yielding a sparse graph.
To this end, I propose a new method called ASTRICS to measure similarity between clusters of HD data points based on local dimensionality reduction and triangulation of critical alpha shapes.
I show that my ASTRICS similarity measure can facilitate both clustering and visualization of HD data by using it in Stage 2 of a three-stage pipeline: Stage~1 = perform an initial clustering of the data by any method; Stage~2 = let graph nodes represent initial clusters instead of individual data points and use ASTRICS to automatically define edges between nodes; Stage~3 = use the graph for further clustering and visualization.
This trades the critical task of choosing a graph neighbourhood size for the easier task of essentially choosing a resolution at which to view the data. The graph and consequently downstream clustering and visualization are then automatically adapted to the chosen resolution.

\newpage




\section{Introduction}

Identifying meaningful clusters of similar objects in high-dimensional (HD) data is an important but challenging problem in a variety of research fields including computational biology, text processing, and image analysis. In molecular biology, for example, flow cytometry (FC) \cite{Picot2012fc} and mass cytometry (CyTOF) \cite{Bandura2009cytof} are high-throughput single-cell technologies that can quantify the abundance of tens of proteins simultaneously in single cells. They are commonly used to investigate phenotypic heterogeneity among cells in tumours \cite{Bendall2012_profilers_guide, Chen2017_immunophenotyping_by_FC, DiGiuseppe2019_FC_immunophenotyping_B/T-LL, Wang2019_fc-immunophenotyping-plasma-cell-neoplasms, Polakova2019_MC_for_immunoprofiling_solid_tumors, deVries2020_cytometry_of_CRC, Maby2020_phenotyping_TILs}, but this demands identification of biologically meaningful clusters of cells in the data. Furthermore, effective presentation of the data requires visualization in three or fewer dimensions.

HD data poses computational challenges collectively referred to as the ``curse of dimensionality'' \cite{Verleysen2005_curse_of_dimensionality, Assent2012_clustering_hd_data, Newell2016_cytof-review}, which cause many clustering algorithms to often perform poorly in HD space. However, their biggest hindrance is not necessarily the actual dimensionality but rather the number of dimensions that are not intrinsically relevant to the data \cite{DURRANT2009385}. Reducing dimensions to a set of relevant dimensions is therefore a general strategy to combat the curse of dimensionality. \textit{Feature selection} 
is the process of identifying a relevant subset of the original variables (i.e.~features, or dimensions) and discarding potentially irrelevant or unhelpful variables. Unfortunately, this can be challenging or subjective. Other dimensionality reduction (DR) methods involve linear or nonlinear transformation of the data space. 

\textit{Feature extraction} is a general term for linear DR methods that construct a reduced number of new features (i.e.~new axes) by forming linear combinations of existing variables.
A widely known linear method for DR is principal component analysis (PCA) \cite{Jolliffe2016_pca_review}. For given data, PCA finds new orthogonal basis vectors (i.e.~axes), called \textit{principle components} (PCs). The first PC is the direction of maximum variance and each successive PC explains as much of the remaining variance as possible. 
To reduce the dimensionality of $d$-dimensional data, the data is projected onto the first $r$ PCs for some user-determined choice of $r<d$.
Usually, DR is applied as a fixed pre-processing step before clustering. Alternatively, Ding et al.~\cite{Ding2002_adaptive_DR} propose integrating adaptive DR into the clustering process by alternating between DR, using PCA, and clustering, which can be performed using, for example, the $K$-means algorithm \cite{Lloyd1982_kmeans}. Their initial DR step applies PCA to all of the data and then all subsequent DR steps use only the centroids of the $K$ clusters in the original $d$-dimensional space. The number of clusters, $K$, and the reduced number of dimensions, $r$, are chosen by the user, though $r=K-1$ is recommended.
A drawback shared by the fixed and adaptive linear DR approaches is the assumption that a single $r$-dimensional subspace is suitable for all of the data. In reality, different dimensions (original or transformed) could be relevant to different subsets of the data. All of the dimensions could ultimately be relevant, even if all clusters are inherently low-dimensional (LD). Subspace clustering algorithms address this by identifying clusters in different subspaces of HD data \cite{Parsons2004_subspace_clustering_review}. But with infinitely many possible subspaces and even $2^d$ axes-parallel subspaces of a $d$-dimensional space, subspace clustering algorithms use heuristics or random projections to remain computationally tractable.

Nonlinear DR methods allow more complex LD embeddings than linear projections. They are preferred for visualizing HD data because of their superior ability to separate clusters in as few as two or three dimensions. A leading nonlinear DR method frequently used for visualizing CyTOF data is t-distributed stochastic neighbour embedding (t-SNE) \cite{vandermaaten2008_tsne}. The LD t-SNE map roughly preserves local neighbourhoods of data points from the HD space but allows distant points to appear further apart in the map. Uniform manifold approximation and projection (UMAP) \cite{mcinnes2020umap} is a newer method competitive with t-SNE that similarly preserves local structure in an LD embedding of the data and has also been adopted for visualization of single-cell data \cite{Becht2019umap-sc}.
Nonetheless, most nonlinear DR methods, including t-SNE, are founded on distances in the HD space, which can be highly sensitive to noise, and are therefore themselves susceptible to the curse of dimensionality. Consequently, linear DR is often still required.
Another limitation of t-SNE is its dependence on an input \textit{perplexity} parameter, which controls the size of the local neighbourhoods that the map tries to preserve, in addition to other hyperparameters. The t-SNE map can artificially fragment clusters if the perplexity is too small, or it can fail to separate clusters if the perplexity is too large. Moreover, a single choice of perplexity might not be suitable for all of the data if the number of points varies widely between true clusters. The same is true for UMAP, which also requires a neighbourhood size parameter to be input. 
Although some clustering algorithms for CyTOF data function on a 2D t-SNE map of the data instead of in the original HD space \cite{Shekhar2014_accense, Becher2014_densvm, Chen2016_clusterX}, in practice this only introduces more parameters, sensitivity, and variability into the clustering process without any tangible benefit except for a satisfying agreement between the visualization and clustering results. Of course, if the visualization is incorrect, then so too will be the clustering results.

Another approach to clustering HD data is to convert the data to a graph and then use a graph-based clustering algorithm. In the graph, nodes typically represent data points and edges represent relationships between nodes. The edges may be unweighted or they may be weighted by some measure of similarity between their adjoining nodes.
PhenoGraph \cite{Levine2015phenograph}, which was originally developed for CyTOF data, is an example of a clustering algorithm that follows this approach. Given a user-specified value of its input parameter $k$, PhenoGraph starts by constructing a $k$-nearest neighbour ($k$NN) graph from the data, whereby each node (data point) is connected to its $k$ nearest neighbouring nodes according to some measure of distance (e.g.~Euclidean distance, which is the default in PhenoGraph). It then refines the graph by weighting the edges based on the proportions of neighbours that are shared between nodes. Specifically, each existing edge between two nodes in the $k$NN graph is weighted by the Jaccard similarity coefficient \cite{Jaccard1912} of their neighbourhoods. The authors call the weighted graph the ``Jaccard graph''. PhenoGraph finally identifies clusters by applying the Louvain method \cite{Blondel2008louvain}, a modularity-based community detection algorithm \cite{Fortunato2010}, to the Jaccard graph. 
Although less susceptible than clustering algorithms that operate directly on HD data points, methods based on $k$NN graphs are not immune to the curse of dimensionality because $k$NN searches depend on measurements of distances between the data points. Therefore, DR may still be beneficial.

Furthermore, setting input parameters presents a challenge for users of most clustering algorithms. Often, the number of clusters $K$ must be specified by the user, which can be difficult because the true number of clusters is usually unknown. In PhenoGraph, the number of clusters in the Jaccard graph is determined automatically by the Louvain method. However, the user-specified parameter $k$ indirectly controls the number of clusters output by PhenoGraph because it controls the connectivity and topology of the graph.
Setting input parameters that control the number of clusters becomes especially challenging when the goal of clustering is to discover novel groups of data points that have unknown existence prior to analysis, such as new cell types in CyTOF data. This challenge could potentially be eliminated if a graph could be constructed from the data in a fully automated manner without user-specified parameters.
As in PhenoGraph, clustering of the graph nodes could subsequently be performed using an algorithm that automatically determines the number of output clusters (e.g.~the Louvain method).
Because a similarity matrix, which encodes the pairwise similarities between objects, can be interpreted as the adjacency matrix of a weighted graph, any parameter-free measure of similarity between nodes would enable parameter-free graph construction. However, to be useful, it would require two important properties. First, in order for clustering to be computationally practical, the similarity of any two distant nodes would have to be 0 (which equates to no edge between them) such that the graph would be sparse. Second, the similarity measure must be suitable for HD data.

Unfortunately, given two data points represented by position vectors in continuous HD space, I am not aware of any measure of similarity between them that would satisfy the two properties above in addition to being parameter-free, and neither could I conceive one. There are various possibilities for a measure of similarity between two data points in continuous space, including the following examples: any decreasing function of the distance between them measured by an $\ell^p$ norm (e.g.~$1/x_{ij}$ or $ e^{-x_{ij}}$, where $x_{ij}$ is the Euclidean distance between data points $i$ and $j$); the cosine of the angle between their position vectors; and the correlation between their position vectors when the vectors are treated as series of values for one-dimensional random variables.
However, without the introduction of an arbitrary parameter or cutoff value, none of these would yield a sparse graph. Additionally, some of these similarity measures can be unsuitable for HD data.
On the other hand, if each graph node were to represent a group of data points instead of an individual data point, then more information, including the distribution of data points within each group, would become available to use for defining the similarity between two graph nodes. In turn, this would increase the potential for defining a parameter-free similarity measure that satisfies the required properties. 

For a similarity measure between groups of data points rather than between individual data points to be useful for a graph-based clustering approach, an initial grouping of the data points to form the graph nodes would be necessary. Some form of fine-grained clustering would therefore have to be performed before parameter-free graph construction could proceed, which seems counterproductive. However, the notion that clustering is difficult due to the curse of dimensionality and due to the challenge of setting input parameters assumes that we seek an optimal partition of the data in the sense of matching some ground truth as closely as possible. Obtaining a suboptimal partition with high purity is much easier (the \textit{purity} of a cluster is the largest proportion of its constituent data points that belong to the same true cluster; in terms more familiar to biologists, it can be equated to \textit{specificity}). A high-purity partition can be achieved by many clustering algorithms if the number of clusters $K$ is chosen to be much larger than the expected true number of clusters. 
Indeed, after comparing 18 clustering algorithms for FC and CyTOF data, Weber and Robinson recommend setting input parameters conservatively (i.e.~$K$ too large or $k$ too small) so as to generate too many clusters and then manually merging clusters where appropriate during interactive downstream analysis \cite{Weber2016}. 
When following their own recommendation, Weber and Robinson found that FlowSOM \cite{VanGassen2015_flowsom} performed well and was the overall best clustering algorithm for FC and CyTOF data, even though it and other algorithms performed poorly when operated in a mode that selected the number of clusters $K$ completely automatically. 
Thus, an initial fine-grained clustering could be obtained using an existing clustering algorithm with the number of clusters chosen to be relatively large. Although this would still require a user-specified input parameter,
it would convert the problem of essentially choosing the final number of clusters to the less-critical task of choosing a resolution at which to represent the data by a graph. Crucially, the exact choice made by the user would become less important because the parameter-free similarity measure would automatically adapt the graph to the chosen resolution.

Motivated by the ideas above, I propose a novel measure of similarity between groups of HD data points based on geometry of alpha shapes \cite{Edelsbrunner1983_alpha-shapes} in locally determined 2D and 3D subspaces. Alpha shapes generalize the convex hull to allow non-convex boundaries to be defined for point sets, and they can have multiple disjoint regions. This makes them appealing for defining cluster boundaries and testing whether a set of points can be rationally segregated into disjoint clusters. My new similarity measure is computed using a method that I call \textit{Alpha Shape TRIangulation in loCal Subspaces} (ASTRICS). ASTRICS has a single parameter, $\alpha$, but geometric considerations are used to automate selection of its value for any pair of clusters.
Importantly, my ASTRICS similarity measure satisfies the three key requirements: it does not have any user-controlled parameters, it has value 0 for well-separated clusters, and it is suitable for HD data.

In addition, I describe a three-stage pipeline built around ASTRICS to cluster and visualize HD data. In the first stage of my pipeline, $N$ data points are partitioned into $K$ \textit{seed clusters} using an existing clustering algorithm, where $K$ is determined by the user and chosen to be considerably larger than the expected true or desired number of clusters. 
This step leverages the ability of many conventional clustering algorithms to generate high-purity clusters when intentionally used to over-cluster HD data, even if they would be unsuitable for finding the true clusters. At this fine-grained level, clustering algorithms that make otherwise questionable assumptions such as convexity of clusters are still applicable. Also, careful optimization of input parameters or the clustering procedure is not necessary at this stage. 
$K$-means \cite{Lloyd1982_kmeans} is a simple and popular clustering algorithm, but it assumes that clusters are convex and is generally unsuited to clustering HD data. Nevertheless, it can still be a suitable choice of algorithm for the first stage of my pipeline, and its implementation is usually very fast. The self-organizing map (SOM) \cite{Kohonen1990_SOM} utilized by FlowSOM is another fast approach potentially suitable for the task.
The second stage of my pipeline uses ASTRICS to compute a $K \times K$ similarity matrix for the $K$ seed clusters, which defines a weighted graph with $K$ nodes. ASTRICS combats the curse of dimensionality here without assuming that any dimensions are universally irrelevant to all of the data by performing DR locally on pairs of seed clusters. In the third and final stage of my pipeline, a community detection algorithm is applied to the ASTRICS similarity matrix/graph to obtain the final set of clusters and the graph is visualized in two or three dimensions using force-directed layout~\cite{Fruchterman1991_force-directed_placement}.

This three-stage strategy for clustering HD data bears resemblance to the SWIFT clustering algorithm \cite{Naim2014_SWIFT_part1}. SWIFT performs an initial over-clustering step using a mixture of Gaussians and then uses Fisher's linear discriminant analysis (LDA) \cite{Fisher1936} to locally project pairs of clusters (i.e.~Gaussian components in the mixture) onto one dimension that best discriminates them. It also projects them onto each of the original data axes and the PCs. The algorithm successively merges clusters if their combined 1D kernel density estimate is unimodal in every one of those projections and if another threshold criterion is met. In contrast, ASTRICS computes an explicit similarity measure that could, for example, be computed for pairs of clusters at the intermediate step of the SWIFT algorithm. This is more versatile than the hard criteria for merging clusters in SWIFT.

Herein, I detail ASTRICS and provide concrete examples of my associated three-stage pipeline for clustering and visualizing HD data. I show that ASTRICS is suitable for measuring similarities between clusters in real-world HD data by applying my pipeline to three very different datasets: publicly available CyTOF data \cite{Samusik2016}, images of handwritten digits from the MNIST database \cite{Lecun1998_mnist}, and text data from the \textit{20 Newsgroups} corpus \cite{20newsgroups_home}.
Quantitatively, my pipeline yields excellent clustering results. Qualitatively, it produces visualizations of HD data that are comparable to those generated by applying t-SNE, a state-of-the-art method, to the centroids of the seed clusters from the first stage of my pipeline. Moreover, my visualization approach may better preserve global relationships than t-SNE. 
The representation of data in my pipeline by a graph, albeit at a coarser resolution than the original data, that is used for both clustering and visualization is somewhat novel as it moves away from the traditional paradigm of using completely separate algorithms for these tasks.
Overall, ASTRICS can enable effective clustering and visualization of HD data using existing algorithms that would otherwise be either unsuitable for or not directly applicable to HD data.
Used in the way that I describe, ASTRICS dilutes the challenge of setting input parameters for clustering by requiring the user to only choose a resolution at which to view the data and allowing everything else to be automated. 
Alternatively, the visualization afforded by ASTRICS could be used to guide selection of the final number of clusters or to select an appropriate resolution from a hierarchical clustering (HC; i.e. nested clusters). 
In sum, I believe that ASTRICS can become a useful tool for analyzing HD data.

\section{Methods}

\subsection{Computing inter-cluster similarities using local DR and alpha shapes}

I developed ASTRICS as a fully automated method to measure the pairwise similarities between clusters in partitioned HD data. The full ASTRICS algorithm starts with two preliminary steps, one to detect outliers and one to reduce the overall computational complexity. These steps are described later. Otherwise, ASTRICS consists of the following two fundamental steps:
\begin{enumerate}
    \item \textbf{The ``CS'' step} (referring to the ``CS'' in ``ASTRICS''): Local DR by projecting individual pairs of clusters onto 2D and 3D subspaces specific to each pair.
    \item \textbf{The ``ASTRI'' step}: Computation of the similarity between two clusters via triangulation of alpha shapes in the 2D and 3D subspaces.
\end{enumerate}
These two fundamental steps are described in detail below for any pair of clusters.

\subsubsection*{The CS step}

The CS step reduces dimensionality in order to combat potential problems associated with the curse of dimensionality while retaining important information about the data. Because different dimensions could be relevant to different clusters, DR is performed locally on pairs of clusters instead of using a common lower-dimensional subspace for all of the data. This ensures that computation of the similarity between any two clusters uses only dimensions that are specifically relevant to those clusters. Also, the consideration of just two clusters at a time makes two or three dimensions sufficient to adequately represent them. In turn, this allows their similarity to be computed and easily visualized using computational geometry.

\begin{figure}[htbp]
    \centering
    \includegraphics[width=\textwidth]{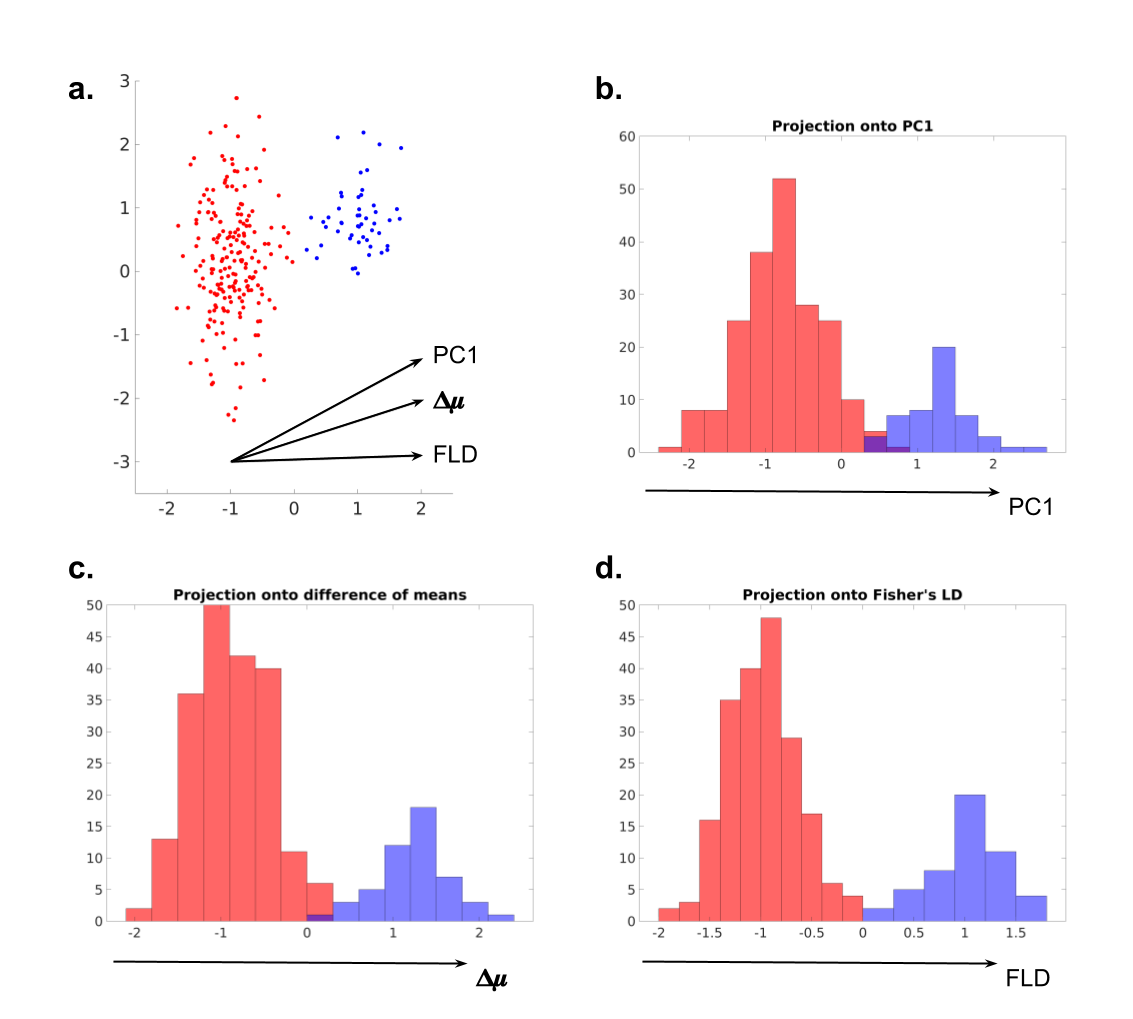}
    \caption[Three possibilities for the first subspace dimension in local DR.]{\textbf{Three possibilities for the first subspace dimension in local DR.} \textbf{(a)} Directions of the first principal component (PC1), the difference of centroids ($\Delta\mu$), and Fisher's linear discriminant (FLD) illustrated for a pair of 2D clusters. \textbf{(b--d)} Histograms of the projections of the two clusters onto the 1D subspaces spanned by PC1 (b), $\Delta\mu$ (c), and FLD (d), which illustrate the increasing resolution of the projected clusters achieved by the three choices in that order for the first subspace dimension in local DR.}
    \label{fig:ASTRICS_DR_methods}
\end{figure}

For the CS step in ASTRICS, three alternative methods were included for performing DR locally on a pair of clusters (\hlcy{Figure}~\ref{fig:ASTRICS_DR_methods}a). 
In the order that they are described below, the three possible methods have increasing ability to resolve the two clusters in a pair of clusters but also increasing susceptibility to overfitting. They also have different computational costs. The choice of method for local DR is optional and may be made based on the user's tolerance for potential overfitting balanced against their tolerance for suboptimal resolution of clusters, as well as on considerations of computational costs.
In all three methods, for a given pair of clusters, we start by eliminating any dimensions (columns of the data matrix) that have fewer than two nonzero elements, since these will be uninformative yet could slow down computations.
The first method then simply proceeds to apply PCA to the pair of clusters and project them onto the subspace spanned by the first two or three PCs (\hlcy{Figure}~\ref{fig:ASTRICS_DR_methods}b). To ensure that both clusters always contribute equally to PCA, the contributions of the data points should be weighted based on the number of points in each cluster.
Local PCA has previously been proposed in clustering and other contexts \cite{Kambhatla1997_local-pca, Lai1999_local-PCA_fMRI, mueller-2003-compression, Huang2008_local-PCA-SOM, ariascastro2013_spectral-clust_local-pca}. Although this is generally quite effective, it does not guarantee that the clusters will be separated in the LD subspace even if they are separated in the HD space. This is because the directions of greatest variance (i.e.~the leading PCs) are not necessarily the ones relevant to the separation of the clusters. PCA would be ineffective at separating two clusters if their separation is very small in relation to their variances along multiple directions.
Ideally, the LD projection of two clusters should retain any separation present in the HD space. This motivates the remaining two methods for local DR that are implemented as options in ASTRICS. 

I refer to the second local DR method as ``Centroids+PCA''. In this method, the first subspace dimension is chosen to be the direction $\vv{v}$ of the straight line through both cluster centroids (\hlcy{Figure}~\ref{fig:ASTRICS_DR_methods}c). This is the same dimension that the adaptive DR method of Ding et al.~\cite{Ding2002_adaptive_DR} would select for $K=2$ clusters. This choice of dimension maximizes the projected separation of the cluster centroids. PCA (again weighted) is then used to find second and third dimensions for the LD subspace. Specifically, the pair of clusters is projected onto the orthogonal complement of $\vv{v}$ (i.e.~the $\vv{v}$ component is subtracted from the data) and then PCA is applied to this orthogonal projection. The resulting first and second PCs become the second and third subspace dimensions, respectively. Subtraction of the $\vv{v}$ component is necessary to ensure that the PCs obtained by PCA are orthogonal to the already determined first subspace dimension. Subsequently, the pair of clusters are projected onto the subspace spanned by $\vv{v}$ and the first one or two PCs that were computed.

Although the Centroids+PCA method maximizes the separation of the cluster centroids in the LD projection, this does not necessarily yield the best separation of the clusters when also considering their variances. Thus, the third method for local DR in ASTRICS, ``LDA+PCA'', uses LDA to determine the first subspace dimension. LDA finds the direction $\vv{u}$ that maximizes the separation of the cluster centroids relative to within-cluster variance in the 1D projection of the clusters onto $\vv{u}$ (\hlcy{Figure}~\ref{fig:ASTRICS_DR_methods}d). The remainder of the LDA+PCA method proceeds in the same way as the Centroids+PCA method (with $\vv{v}$ replaced by $\vv{u}$).

However, LDA is known to have problems in high dimensions \cite{Qiao2008_LDA-HD/LSS, Mai2013_review_LDA_HD, Sharma2015_LDA_SSS_overview}, especially if the number of dimensions is greater than the number of data points. For instance, LDA usually involves inversion of a pooled within-cluster covariance matrix, but this could be singular and therefore not invertible. Another severe problem arises from `small' sample sizes, our notion of which must be adjusted in high dimensions. Because hypervolume increases exponentially with the number of dimensions, sample size needs to be exponentially larger than the number of dimensions in order to avoid `small sample' effects. Otherwise, LDA can overfit the noise in the data. 
To a lesser but still substantial degree, this can also occur when projecting two such sets of HD points onto the direction $\vv{v}$ through both of their centroids. Thus, the Centroids+PCA method is also afflicted by this `small sample size' problem.

Since ASTRICS performs DR locally on pairs of clusters, the `small sample size' problem could be prevalent. To address the problem, both the Centroids+PCA and LDA+PCA methods are modified by introducing two additional steps at the start for every pair of clusters.
First, if the number of data points in a cluster pair is less than $2^d$, we locally apply just PCA or, for sparse data, singular value decomposition (SVD, using the function \texttt{svds} in MATLAB, MathWorks) to initially reduce the number of dimensions from $d$ to the largest integer $r$ such that the number of data points in the cluster pair is at least $2^r$.
Second, Hotelling's T-squared test for two multivariate samples is performed to test whether the cluster means are significantly different at the 5\% level. If they are, then we proceed with the Centroids+PCA or LDA+PCA method as already described. Otherwise, we revert to using the PCA-only method instead because it would be inappropriate to optimize separation of the cluster means when this is not statistically significant. 

At face value, the PCA-only method certainly has the lowest computational cost of the three methods. On the other hand, the Centroids+PCA and LDA+PCA methods, unlike PCA, guarantee that separation of the clusters is factored into the choice of subspace. Moreover, the superior ability of Centroids+PCA and LDA+PCA to separate clusters in the projection can surprisingly reduce the overall time complexity of ASTRICS by reducing the number of computations required in the ASTRI step. Below, I introduce some notation and then outline the step-by-step algorithms for the three possible local DR methods for any pair of clusters. Later, I compare their results for real CyTOF data, MNIST images of handwritten digits, and newsgroups text documents. Overall, all three local DR methods provide good results, but I recommend LDA+PCA.

\subsubsection*{Notation}
Let $\setX_i$ and $\setX_j$ denote the sets of data points $x\in\mathbb{R}^d$ belonging to the clusters with labels $C_i$ and $C_j$ respectively, and let $\setX_{ij} = \setX_i \cup \setX_j$. Denote the cardinality (i.e.~size) of a set $\setX$ by $|\setX|$. Without loss of generality, always choose the cluster indices $i$ and $j$ such that $|\setX_i|\geq|\setX_j|$, i.e.~cluster $C_j$ is always the one with the fewest data points in a given pair of clusters.
Also, let $\proj_W(\setX)$ denote the orthogonal projection of a set of data points $\setX$ onto a subspace $W \subset \mathbb{R}^d$ and let $W^\perp$ denote the orthogonal complement of $W$. I use $W_{ij}$ to denote the final LD subspace, determined by local DR, onto which the data points $\setX_{ij}$ will be projected in order to subsequently measure the similarity of clusters $C_i$ and $C_j$. With this notation, also now define $\setY_{ij}=\proj_{W_{ij}}(\setX_{ij})$ and $\setY_\beta^{ij}=\proj_{W_{ij}}(\setX_\beta)$ ($\beta\in\{i,j\}$). In words, $\setY_i^{ij}$ and $\setY_j^{ij}$ are the projections of the data points belonging to, respectively, cluster $C_i$ and cluster $C_j$ onto the LD subspace $W_{ij}$ determined by performing local DR on that pair of clusters. The set $\setY_{ij}$ is the projection onto $W_{ij}$ of the set of all data points belonging to either cluster $C_i$ or cluster $C_j$. Note that $\setY_{ij} = \setY_i^{ij} \cup \setY_j^{ij}$. Furthermore, I assume that vectors are column vectors, and $\vv{v}^t$ will be used to denote the transpose of any vector $\vv{v}$. Thus, by assumption, $\vv{v}$ is a column vector and $\vv{v}^t$ is a row vector. The Euclidean ($\ell^2$) norm of any vector $\vv{v}$ will be denoted by $\norm{\vv{v}}_2$. 
The notation $\floor{a}$ denotes the largest integer that is not greater than the scalar $a$.

\subsubsection*{PCA-only algorithm for local DR}
For a given pair of clusters, $C_i$ and $C_j$, do the following:
\begin{enumerate}
    \item Set $r$ to be the desired reduced number of dimensions ($r\in\{2,3\}$).
    \item Assign a weight of 1 to each data point in $\setX_i$. Assign a weight of $|\setX_i|/|\setX_j|$ to each data point in $\setX_j$.
    \item Compute the first $r$ PCs of $\setX_{ij}$ using PCA with observation weights as specified in step 2. Denote the set of these $r$ PCs by $\setPC^r(\setX_{ij})$.
    \item Define the basis $\setB_{W_{ij}}^r=\setPC^r(\setX_{ij})$ and the subspace $W_{ij} = \spanset(\setB_{W_{ij}}^r)$.
    \item Compute $\setY_i^{ij} = \proj_{W_{ij}}(\setX_i)$ and $\setY_j^{ij} = \proj_{W_{ij}}(\setX_j)$, both expressed in the basis $\setB_{W_{ij}}^r$.
    \item Output $\setY_i^{ij}$ and $\setY_j^{ij}$.
\end{enumerate}

\subsubsection*{Centroids+PCA and LDA+PCA algorithms for local DR}
For a given pair of clusters, $C_i$ and $C_j$, do the following:
\begin{enumerate}
    \item Set $r$ to be the desired reduced number of dimensions ($r\in\{2,3\}$).
    \item Assign a weight of 1 to each data point in $\setX_i$. Assign a weight of $|\setX_i|/|\setX_j|$ to each data point in $\setX_j$.
    \item If $|\setX_{ij}| \geq 2^d$, define $\setZ_i=\setX_i$ and $\setZ_j=\setX_j$ and go to step 5. Else, set $\delta = \floor{\log_2\left(|\setX_{ij}|\right)}$ and compute the first $\delta$ PCs or, for sparse data, right-singular vectors of $\setX_{ij}$ using PCA or SVD with observation weights as specified in step 2. Denote the set of these $\delta$ PCs or right-singular vectors by $\setV^\delta(\setX_{ij})$.
    \item Define the subspace $\Omega=\spanset(\setV^\delta(\setX_{ij}))$. Compute $\setZ_i=\proj_\Omega(\setX_i)$ and $\setZ_j=\proj_\Omega(\setX_j)$, both expressed in the basis $\setV^\delta(\setX_{ij})$ for $\Omega$.
    \item Compute $\vv{v}=\vv{\mu_i}-\vv{\mu_j}$, where
    \[\vv{\mu_\beta} = \dfrac{1}{|\setX_\beta|}\sum_{z\in\setZ_\beta}z\]
    is the centroid (mean) of $\setZ_\beta$ ($\beta\in\{i,j\}$).
    \item Compute the $\rho \times \rho$ within-cluster covariance matrices $\Sigma_i=\covar(\setZ_i)$ and $\Sigma_j=\covar(\setZ_j)$, where $\rho=\min\{d,\,\delta\}$.
    \item Compute the pooled within-cluster covariance matrix 
    \[\Sigma_\mathrm{pooled} = \dfrac{(|\setX_i|-1)\Sigma_i + (|\setX_j|-1)\Sigma_j}{|\setX_i|+|\setX_j|-2}.\]
    \item Assuming that $\Sigma_\mathrm{pooled}$ is invertible, compute $\vv{\omega}={\Sigma_\mathrm{pooled}}^{-1}\vv{v}$.
    \item Compute Hotelling's T-squared statistic
    \[T^2 = \dfrac{|\setX_i|\,|\setX_j|}{|\setX_i|+|\setX_j|}\,\vv{v}^t\,\vv{\omega}.\]
    \item Convert $T^2$ into the $F$-statistic
    \[F = \dfrac{(|\setX_i|+|\setX_j|-\rho-1)}{(|\setX_i|+|\setX_j|-2)\rho}\,T^2.\]
    \item Set $F_*$ equal to the 95\% confidence level of the $F$-distribution with first parameter $\rho$ and second parameter $(|\setX_i|+|\setX_j|-\rho)$.
    \item If $F<F_*$, stop here and revert to the PCA-only algorithm (note that the PCs do not need to be recomputed if they were already computed in step 3). Else, determine the direction $\vv{u}=\vv{v}/\norm{\vv{v}}_2$ if using Centroids+PCA or $\vv{u}=\vv{\omega}/\norm{\vv{\omega}}_2$ if using LDA+PCA.
    \item Define $U=\spanset\{\vv{u}\}$ and compute $\Phi=\proj_{U^\perp}(\setZ_i \cup \setZ_j)$.
    \item Compute the first $(r-1)$ PCs of $\Phi$ using PCA with the observation weights specified in step 2. Denote the set of these $(r-1)$ PCs by $\setPC^{r-1}(\Phi)$.
    \item Define the basis $\setB_{W_{ij}}^r=\{\vv{u}\} \cup \setPC^{r-1}(\Phi)$ and the subspace $W_{ij} = \spanset(\setB_{W_{ij}}^r)$.
    \item Compute $\setY_i^{ij} = \proj_{W_{ij}}(\setZ_i)$ and $\setY_j^{ij} = \proj_{W_{ij}}(\setZ_j)$, both expressed in the basis $\setB_{W_{ij}}^r$.
    \item Output $\setY_i^{ij}$ and $\setY_j^{ij}$.
\end{enumerate}

\subsubsection*{The ASTRI step}

The ASTRI step measures the similarity between two clusters after the CS step has reduced their dimensionality. To this end, I sought a suitable measure of similarity between two sets of $r$-dimensional points $\setY_i^{ij}$ and $\setY_j^{ij}$.

Intuitively, similarity should be inversely related to distance. Thus, inter-cluster similarity could be defined using a decreasing function of some measure of distance between clusters. Agglomerative HC algorithms provide several possible distance measures: single-linkage \cite{Gower1969_single-linkage} measures the shortest distance between two points in different clusters; complete-linkage \cite{Krznaric1998_complete-linkage} measures the greatest distance; average-linkage, which has weighted and unweighted variants \cite{RamosEmmendorfer2021_average-linkage}, measures the the average of all pairwise distances between points in different clusters; centroid-linkage \cite{Lance1967_centroid-linkage} measures the distance between cluster centroids. 
However, these distance measures behave in undesirable ways when cluster size is varied, and they lack suitability when clusters are overlapping. Those that require computation of pairwise distances between many points can also be computationally intensive. Moreover, the choice of function to map distance to similarity is arbitrary, and, without the further introduction of an arbitrary threshold, any reasonable choice of function would lead to a full similarity matrix in which all pairs of clusters have nonzero similarity. This is undesirable because a sparse similarity matrix would greatly benefit the computational cost of algorithms that use it as input and also benefit our interpretation of it as a graph. 

Alternatively, I considered using the Bhattacharyya coefficient to measure inter-cluster similarity. This is a statistical measure of similarity, or overlap, between two distributions, and it has previously been applied in data mining \cite{Aherne1997_Bhattacharyya-metric-freq-coded-data, BenAyed2015_distribution-matching-with-Bhattacharyya-similarity, Patra2015_Bhattacharyya-coeff-for-CF}. Unlike the popular Kullback-Leibler divergence \cite{Kullback1951_KL-divergence} measure of dissimilarity between two distributions, the Bhattacharyya coefficient is symmetric, a desirable property. However, despite being an appealing measure of similarity between two distributions of points and having a simple geometric interpretation \cite{Aherne1997_Bhattacharyya-metric-freq-coded-data, BenAyed2015_distribution-matching-with-Bhattacharyya-similarity}, it has a major limitation that prevents it from being used to measure inter-cluster similarity: the domain of the clusters needs to be discretized and the Bhattacharyya coefficient is sensitive to the grid element sizes. Unfortunately, there is no obvious method to automatically choose a suitable discretization, especially in $>1$ dimension, so the Bhattacharyya coefficient would be sensitive to arbitrary parameters that define the grid.

\begin{figure}[htbp]
    \centering
    \includegraphics[width=\textwidth]{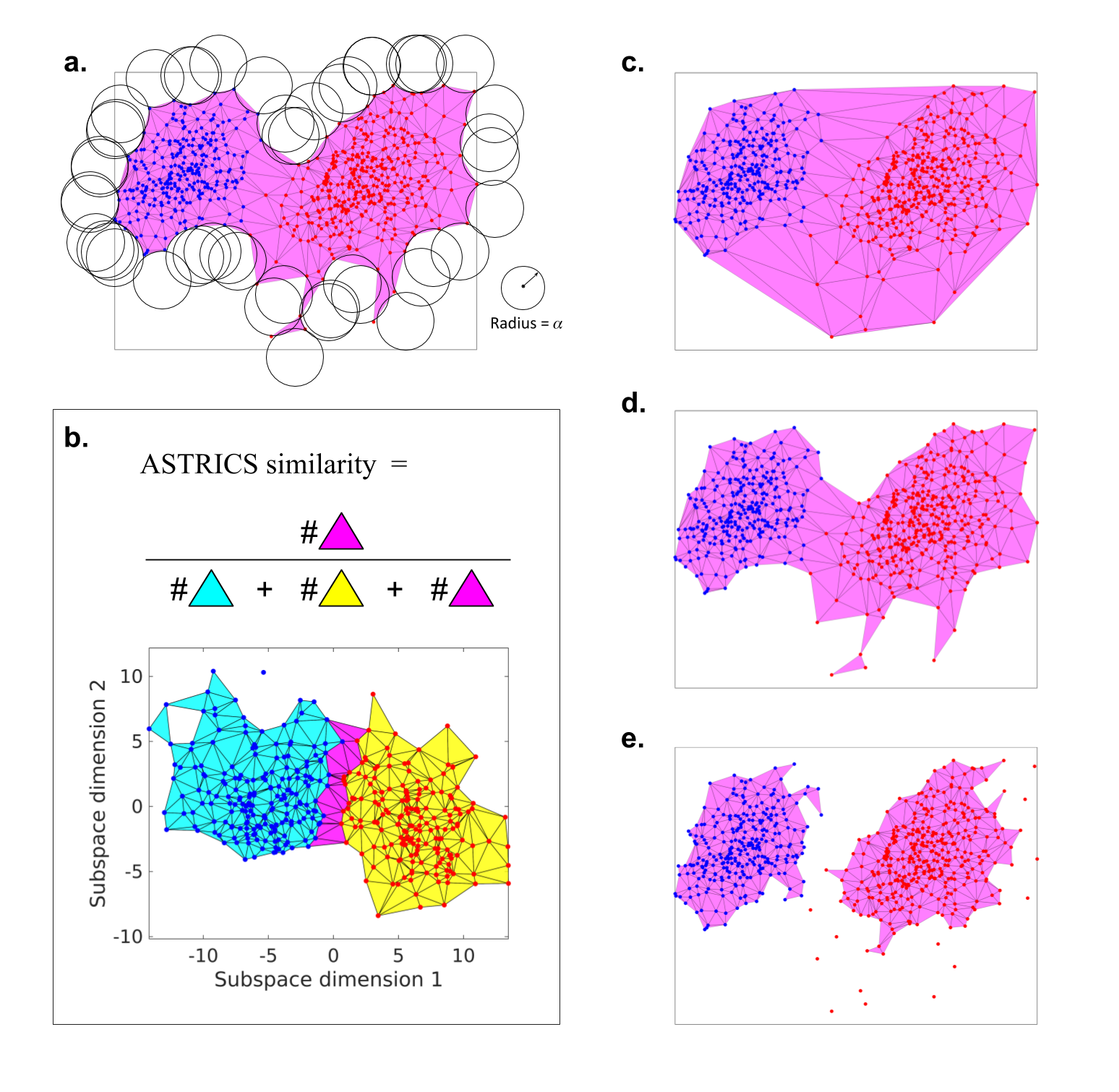}
    \caption[Computing inter-cluster similarity by triangulating an automated choice of alpha shape.]{\textbf{Computing inter-cluster similarity by triangulating an automated choice of alpha shape.} \textbf{(a)} Construction of an alpha shape. \textbf{(b)} Graphical illustration of the ASTRICS similarity for a pair of clusters (blue points and red points) that have been projected onto a 2D subspace by local DR. The colours and points bear no relation to the colours or points in any other figure panels. \textbf{(c)} Convex hull of the union of two sets of points (blue and red clusters). \textbf{(d)} Smallest alpha shape enclosing all points in the union of the two point sets. \textbf{(e)} Smallest alpha shape enclosing all points from at least one of the two point sets. In this example, the inter-cluster similarity measured by ASTRICS is 0 for the blue and red clusters of points.}
    \label{fig:ASTRICS_computation}
\end{figure}

Instead, I propose a new measure of similarity based on alpha shapes and triangulation. Alpha shapes provide a one-parameter generalization of the convex hull that enables non-convex boundaries to be defined for point sets (\hlcy{Figure}~\ref{fig:ASTRICS_computation}a). Triangulation allows the domain of a set of points to be discretized without the need for any arbitrary parameters. Note, however, that a triangulation is not a suitable discretization to use for the Bhattacharyya coefficient because it would be unclear how to bin the points, which form the vertices of the mesh elements. For two sets of points in two or three dimensions, an alpha shape is computed for their union. As I discuss later, an appropriate critical value is used to automate selection of the one parameter, $\alpha$, needed to define the alpha shape. This removes any need for parameters to be specified by the user. Next, triangulation of the points generates a discrete mesh that fills the alpha shape. Finally, the similarity of the two point sets is defined to be the fraction of triangulation mesh elements that have at least one vertex from both point sets (\hlcy{Figure}~\ref{fig:ASTRICS_computation}b). I call these mesh elements ``multicolour'' elements.

My similarity measure has attractive properties. Firstly, it is bounded between 0 and 1. It has a value of 0 only when the alpha shape has disjoint regions that each exclusively enclose points belonging to just one point set. Thus, for suitably chosen $\alpha$, two point sets have a similarity of 0 only if they appear to be geometrically distinct from each other. My similarity measure achieves its maximum value of 1 only if every triangulation mesh element has vertices from both point sets, which is only possible if the two sets of points are well mixed. 

Another attractive property of my similarity measure is the full automation of its computation, including selection of its only parameter, $\alpha$. On the other hand, the parameter $\alpha$ provides flexibility for users who choose not to automate its selection. Furthermore, despite the continuum of possible values for $\alpha$, there exist only finitely many unique alpha shapes, corresponding to disjoint intervals of $\alpha$ values, for any finite set of points. Thus, similarity could be computed as a piecewise constant function of $\alpha$ instead of as a single value. However, this would be computationally intensive, and it would not be clear how to use this piecewise function for visualization or clustering.

Finally, my similarity measure is, in fact, a Jaccard similarity coefficient \cite{Jaccard1912} when viewed in the following way. For the point sets $\setY^{ij}_i$ and $\setY^{ij}_j$, let $\setT^{ij}_i$ and $\setT^{ij}_j$ be the sets of triangulation mesh elements with at least one vertex belonging to $\setY^{ij}_i$ and $\setY^{ij}_j$ respectively. Then my similarity measure is the Jaccard similarity coefficient of the sets $\setT^{ij}_i$ and $\setT^{ij}_j$, which is defined to be the ratio of the size of their intersection to the size of their union.

Below, I outline the full algorithm for the ASTRI step of ASTRICS to compute inter-cluster similarities. My similarity measure, which I call the \textit{ASTRICS similarity}, could be computed in two ($r=2$) or three ($r=3$) dimensions. However, for HD data to which ASTRICS is applied, it is difficult to justify a choice of $r=2$ or $r=3$ \textit{a priori}. I argue that the number of dimensions, $r$, that makes it easier to geometrically discriminate two clusters should be chosen. Hence, ASTRICS is performed for $r=2$ and $r=3$ and, for each pair of clusters, the minimum of the two computed similarity scores is taken.

\subsubsection*{ASTRI algorithm for measuring inter-cluster similarity}
For a given pair of clusters, $C_i$ and $C_j$, do the following:
\begin{enumerate}
    \item For an automatically selected value of $\alpha$ and a number of dimensions $r\in\{2,3\}$, compute the $\alpha$-shape for the $r$-dimensional projection $\setY_{ij} = \setY_i^{ij} \cup \setY_j^{ij}$ of the data points $\setX_{ij}$ belonging to the pair of clusters. The automatic selection of $\alpha$ will be discussed later.
    \item Generate a triangulation $\setT^{ij}$ of the points in $\setY_{ij}$ that exactly defines the domain of the $\alpha$-shape. The specific choice of triangulation will be discussed later. The elements of the set $\setT^{ij}$ are $(r+1)$-tuples specifying the vertices (which are elements of $\setY_{ij}$) of triangles in 2D or tetrahedra in 3D.
    \item For $\beta\in\{i,j\}$, identify the set $\setT^{ij}_\beta$ of elements in the triangulation $\setT^{ij}$ that have at least one vertex from the LD projection $\setY_\beta^{ij}$ of the data points $\setX_\beta$ from cluster $C_\beta$.
    \item Identify the set $\setMC_{ij} = \setT^{ij}_i \cap \setT^{ij}_j$ of ``multicolour'' triangulation elements. 
    \item Determine $|\setT^{ij}|$, the total number of elements in the triangulation $\setT^{ij}$.
    \item Determine $|\setMC_{ij}|$, the number of multicolour elements in the triangulation $\setT^{ij}$.
    \item Compute \[A^r_\alpha(C_i,C_j) = \dfrac{|\setMC_{ij}|}{|\setT^{ij}|},\] the fraction of triangulation elements that are multicolour. I call this the $(r,\alpha)$-similarity of clusters $C_i$ and $C_j$ for reduced number of dimensions $r\in\{2,3\}$ and an alpha shape specified by $\alpha$. Let the notation $\alpha=*$ indicate that the value of $\alpha$ was determined automatically for a given pair of clusters.
    \item Compute $A^r_*(C_i,C_j)$ for both $r=2$ and $r=3$.
    \item Set $AS(C_i,C_j)$ to be the minimum of the two $(r,*)$-similarity scores computed above for the pair of clusters, i.e.~\[AS(C_i,C_j) = \min\{A^r_*(C_i,C_j) : r\in\{2,3\}\}.\]
    \item Output $AS(C_i,C_j)$. This is the new inter-cluster similarity measure that I propose for HD data, which I call the \textit{ASTRICS similarity}.
\end{enumerate}


\subsection{Automated selection of \texorpdfstring{$\boldsymbol{\alpha}$}{alpha}}

The parameter $\alpha\in\mathbb{R}$ determines the alpha shape for a set of points $\setY_{ij} \in \mathbb{R}^r$ ($r\in\{2,3\}$). I automate its selection by considering critical alpha shapes. Here, I discuss three possible critical alpha shapes for this purpose. I then explain my choice of critical alpha shape for ASTRICS.

\begin{enumerate}
    \item \textbf{The convex hull (\hlcy{Figure}~\ref{fig:ASTRICS_computation}c).} 
    This corresponds to $\alpha$ values in a disjoint interval containing $\infty$. The convex hull is the most well known example of an alpha shape. A triangulation that fills the convex hull of a set of points is their Delaunay triangulation, another well known concept. Nonetheless, this is a poor choice for ASTRICS. No matter how greatly separated $\setY_i^{ij}$ and $\setY_j^{ij}$ are from each other, the Delaunay triangulation of their union $\setY_{ij}$ must necessarily include multicolour elements. Hence, the ASTRICS similarity matrix would be full because all pairs of clusters would have nonzero similarity regardless of how well separated they are, which is undesirable. Another problem with using the convex hull is that $AS(C_i,C_j)$ would not necessarily decrease as the separation of $\setY_i^{ij}$ and $\setY_j^{ij}$ increases.
    \item \textbf{The smallest alpha shape enclosing all points from $\boldsymbol{\setY_{ij}}$ (\hlcy{Figure}~\ref{fig:ASTRICS_computation}d).} 
    Importantly, the alpha shape can be fragmented into disjoint regions. If there exists an alpha shape that encloses all of the points in $\setY_{ij}$ while completely segregating members of $\setY_i^{ij}$ and $\setY_j^{ij}$ from each other in disjoint regions, then the separation between $\setY_i^{ij}$ and $\setY_j^{ij}$ must be greater than the separation between points in the most sparse region of either $\setY_i^{ij}$ or $\setY_j^{ij}$. This would indicate that $\setY_i^{ij}$ and $\setY_j^{ij}$ are well separated and therefore have zero similarity. Conversely, two point sets would have some nonzero similarity if their separation is not greater than the separation of the most sparsely distributed points.
    \item \textbf{The smallest alpha shape enclosing all points from either $\boldsymbol{\setY_i^{ij}}$ or $\boldsymbol{\setY_j^{ij}}$ (\hlcy{Figure}~\ref{fig:ASTRICS_computation}e).} 
    In this case, we allow points from one of the two point sets to become isolated as long as we enclose all of the points from the other set. Whereas the previous choice of alpha shape is determined by the most sparse region of points in either $\setY_i^{ij}$ or $\setY_j^{ij}$, this choice is determined by the most sparse region of just one of the two point sets, specifically the one whose most sparse region is more dense than the other's. The similarity of two sets of points will be zero if it is possible to enclose all of the members of one set within one or more regions of an alpha shape without enclosing any members of the other set in those same regions.
\end{enumerate}

Options 2 and 3 are both justifiable choices. As long as one or the other is used consistently, I suggest that either can be used in ASTRICS depending on a user's preference. Alternatively, any analysis using ASTRICS could be repeated using both options for the automated selection of $\alpha$. This could serve as a means to validate results. Nevertheless, for the following reasons, I prefer Option 3 and therefore use this in ASTRICS.

Firstly, Option~3 has the greatest chance of yielding zero similarity for a pair of clusters and therefore produces the most sparse similarity matrix, which is desirable. Concurrently, this choice of alpha shape may strengthen the similarity of substantially overlapping clusters. For two overlapping sets of points, each of roughly uniform density, the overall local density of points will be greatest in the overlapping region. Consequently, we would expect shrinking the alpha shape to isolate points in a non-overlapping region at a faster rate than it would isolate points in the overlapping region. This would likely result in the overlapping region contributing a greater amount to the triangulation for Option~3 than Option~2. Finally, I argue that, if asked to decide whether two sets of points in fact belong to the same cluster, our intuition would be to base our decision on the more dense points as in Option~3, rather than the most sparse points as in Option~2. After all, our intuition is to identify clusters by searching for increases in density.

\subsection{The choice of triangulation}

Note that the triangulation that defines the domain of an alpha shape is generally not unique for any set of $>(r+1)$ points in $r$ dimensions. To make the triangulation unique in most cases, we can simply demand that it is a subset of the Delaunay triangulation, which is almost surely unique for continuous data. There are exceptions where the Delaunay triangulation is not unique. These occur when a circle can be drawn that exactly passes through four or more data points in 2D, or when a sphere can be drawn that exactly passes through five or more points in 3D. For real, continuous data with independent data points, however, these exceptions have zero probability of occurring. Similarly, exceptions where no triangulation exists occur when all points lie on a straight line in 2D or a plane in 3D.

In my own implementation of ASTRICS in MATLAB, I use the default triangulation provided by the function \texttt{alphaTriangulation}. Unfortunately, the code for this function is proprietary and the description of the function does not provide any details about how the triangulation is computed. Nevertheless, I believe that this function always uses the triangulation that is a subset of the Delaunay triangulation.

\subsection{Preliminary step One: Detection and exclusion of outliers}

As noted earlier, ASTRICS is preceded by two preliminary steps, which are now described. A drawback of using critical alpha shapes in the fully automated ASTRICS algorithm is their sensitivity to outliers. The presence of an outlier in a set of points can have a large effect on the smallest alpha shape that encloses all of the points in the set. Therefore, as a preliminary step, outliers are identified and excluded from all subsequent ASTRICS computations. To do this, we search for outliers within each seed cluster locally using the following algorithm. Note that outlier detection is only performed on seed clusters containing at least 8 points. 

For each seed cluster $C_i$ in isolation, do the following:
\begin{enumerate}
    \item Perform PCA and project the data onto the subspace spanned by the first three PCs to reduce the dimensionality. I chose three dimensions because ASTRICS will be performed in no more than three dimensions; points that are outliers in higher dimensions but not in three dimensions are of no concern.
    \item Standardize the data along each of the three new dimensions using robust estimates of average and scale. Specifically, for each dimension independently, subtract the midmean (i.e.~the mean of the middle 50\%) of the data and divide by the median absolute deviation.
    \item Compute the distance of every point from the origin, which is now the estimated centroid of the seed cluster. Let $\setD_0^i$ denote the distribution of these distances for seed cluster $C_i$.
    \item Also compute the distance of every point from its nearest neighbouring point in the same seed cluster. Let $\setD_\mathrm{nn}^i$ denote the distribution of these distances for seed cluster $C_i$.
    \item Square-root transform all of the computed distances to make the distributions $\setD_0^i$ and $\setD_\mathrm{nn}^i$ become more normal.
    \item Finally, classify points as outliers if they satisfy
    \[\sqrt{D} > Q_3 + \dfrac{3}{2} \times IQR\]
    in both square-root transformed distributions, where $\sqrt{D}$ is the square-root transformed distance and $Q_3$ and $IQR$ are, respectively, the upper quartile and interquartile range of the appropriate square-root transformed distribution.
\end{enumerate}

This search for and subsequent exclusion of outliers prevents ASTRICS from selecting unsuitably large values of $\alpha$ in order to enclose anomalous, distant points in clusters. In turn, this prevents ASTRICS from yielding nonzero similarity for pairs of seemingly well separated clusters on the basis of distant outliers in the clusters.

\subsection{Preliminary step Two: Fast, approximate search for zero-similarity pairs of clusters}

For a dataset partitioned into $K$ seed clusters, there are $(K-1)K/2$ pairs of clusters. Due to the computations of PCs and alpha shapes, using ASTRICS to compute the similarity of every pair of clusters could be computationally intensive for large datasets. Usually, however, some seed clusters will clearly be very distant from other seed clusters in a partitioned dataset with large $K$. Such distant pairs of clusters are likely to have an ASTRICS similarity of 0 when using the automated choice of alpha shape. Hence, when $K\gg2$, a fast preliminary search is performed to find pairs of clusters that can reasonably be assigned a similarity of 0 without applying ASTRICS.

First, PCA is applied to the $K$ seed cluster centroids to change basis from the original $d$ dimensions to a basis composed of all $\min\{d,K-1\}$ PCs so that the seed clusters are spread out as much as possible along each of the new basis dimensions. All of the data are then projected onto each of the PCs, generating $\min\{d,K-1\}$ 1D projections. Subsequently, a similarity of 0 is assigned to any pair of seed clusters whose ranges are disjoint in any of the 1D projections onto the PCs. We then only proceed to apply ASTRICS to pairs of seed clusters that have overlapping ranges in all of the $\min\{d,K-1\}$ 1D projections.

My rationale for this approach is as follows. If a pair of seed clusters can be completely separated in a 1D projection using a set of globally determined, and therefore unlikely to be locally optimal, axes, then those clusters are likely to be even more greatly separated in a subspace determined locally as per the CS step of ASTRICS. As such, I suggest that it would be computationally wasteful to apply ASTRICS to that pair of clusters since their ASTRICS similarity is likely to be 0 or approximately 0. By choosing to assign a similarity of 0 to such pairs of clusters, we lower the number of pairs of clusters to which ASTRICS will be applied, thereby reducing the overall computational cost.

\subsection{A pipeline for visualizing and clustering HD data using ASTRICS}

To test and demonstrate the utility of ASTRICS, I built it into a pipeline for visualizing or clustering HD data. The full pipeline is depicted in general terms in \hlcy{Figure}~\ref{fig:ASTRICS_clustering_pipeline}. This pipeline starts by partitioning the data into $K$ seed clusters (\hlcy{Figure}~\ref{fig:ASTRICS_clustering_pipeline}a) by any method, where $K$ is chosen to be much larger than the expected or desired number of clusters. The exact value of $K$ is not important and therefore does not require optimization. Nonetheless, seed clusters should contain at least four points so that they intrinsically occupy at least three dimensions, but should ideally contain at least $2^3=8$ points to prevent small sample size artifacts. 
In the second step of the pipeline, ASTRICS is used to compute similarities between seed clusters (\hlcy{Figure}~\ref{fig:ASTRICS_clustering_pipeline}b--d) to yield a $K \times K$ similarity matrix, which can be interpreted as a weighted graph (\hlcy{Figure}~\ref{fig:ASTRICS_clustering_pipeline}e). 
In the third and final step of the pipeline, the similarity matrix/graph generated by ASTRICS is used as input to other methods for visualization and further clustering (\hlcy{Figure}~\ref{fig:ASTRICS_clustering_pipeline}f).
For convenience and brevity, a general pipeline of this form will henceforth be referred to as a seed--ASTRICS--clu/vis pipeline. Here, ``seed'' refers to any clustering algorithm (e.g.~$K$-means) used to perform the initial seed clustering of data points, ``clu'' refers to any clustering algorithm that can operate on the similarity matrix/graph induced by applying ASTRICS to the seed clusters (e.g.~a community detection or spectral clustering algorithm), and ``vis'' refers to any method for visualizing a weighted graph (e.g.~force-directed layout).

\begin{figure}[!htbp]
    \centering
    \includegraphics[width=0.7\textwidth]{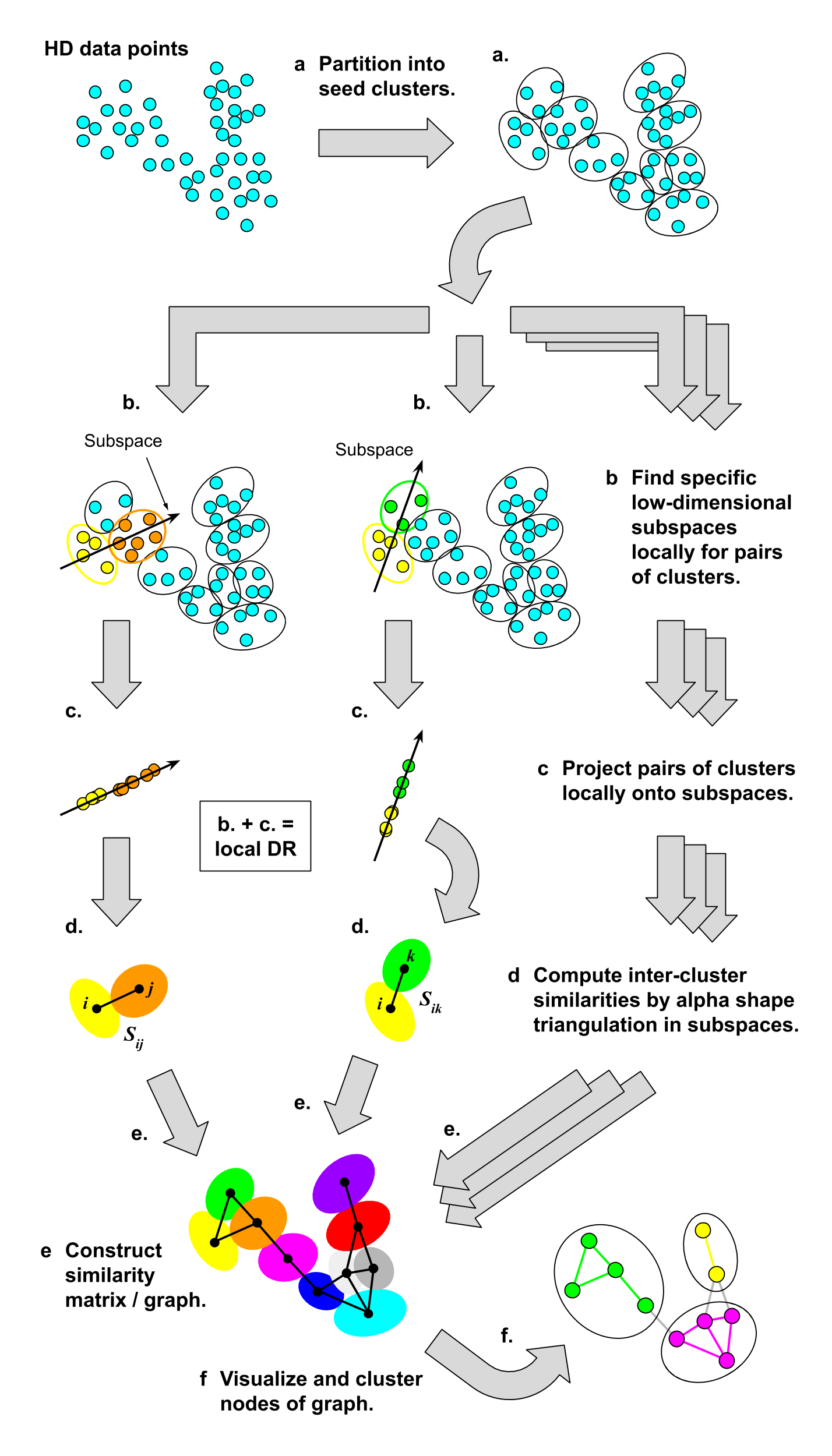}
    \caption[A pipeline for visualizing and clustering HD data using ASTRICS.]{\textbf{A pipeline for visualizing and clustering HD data using ASTRICS.} 
    Caption continues~\dots}
    \label{fig:ASTRICS_clustering_pipeline}
\end{figure}

\begin{figure}[!tbp]
    \centering
    \contcaption{\textbf{(a)} Partition the HD data into $K$ `seed clusters' using any clustering algorithm, e.g.~$K$-means. $K$ should be large. \textbf{(b--d)} Compute the ASTRICS similarity of each pair of clusters: project each pair of clusters onto a low-dimensional subspace using local DR (b--c), then compute the inter-cluster similarity of each pair of clusters using alpha shape triangulation in the subspace (d). \textbf{(e)} Construct a weighted graph with $K$ nodes representing the $K$ seed clusters and edge weights equal to the inter-cluster similarity scores measured by ASTRICS. \textbf{(f)} Visualize the graph using force-directed layout and cluster the nodes using community detection or spectral clustering.}
\end{figure}

For the seed clustering step, I employed either $K$-means or the SOM step from FlowSOM. To ensure that each seed cluster contained at least a specified minimum number of points, chosen to be 8, clusters were merged, when necessary, in a manner similar to centroid linkage. Centroid linkage sequentially merges the two clusters with the shortest distance between their centroids, updating cluster centroids after each merge. My approach computes all of the seed cluster centroids and finds all seed clusters containing the fewest data points. It then merges each one with the seed cluster whose centroid is nearest to its own, before updating the centroids all together after completing all merges. This process repeats iteratively until all seed clusters contain at least the specified minimum number of points. Throughout the seed clustering step, one of the $\ell^2$ (Euclidean) norm, $\ell^1$ (Manhattan, or city block) norm, or correlation distance (i.e.~one minus the Pearson correlation between two vectors) was maintained as the distance measure. I then applied ASTRICS to the seed clusters using the LDA+PCA method for local DR.

Subsequently, the ASTRICS similarity matrix for the seed clusters was treated as a weighted graph and force-directed layout \cite{Fruchterman1991_force-directed_placement} was used to view the graph. Each node in the graph represents a seed cluster and the weight of the edge between two nodes is the ASTRICS similarity of the corresponding seed clusters. The force-directed layout attempts to make edge lengths inversely proportional to the (nonzero) edge weights in the graph so that the greater the similarity of two seed clusters, the closer together their nodes will be in the visualization. Thus, distances in the visualization are, in fact, meaningful, which is generally not the case in a t-SNE map. I found that the force-directed layout algorithm implemented in MATLAB for viewing graphs is sensitive to scaling of the edge weights. For this reason, all of the edge weights were rescaled before visualizing the graph such that the mean nonzero edge weight was 1 after rescaling. This generally resulted in effective visualization where nodes did not concentrate together to the point of being unrecognizable from each other.

To demonstrate clustering of the nodes, I used the Louvain method \cite{Blondel2008louvain}, a modularity-based community detection algorithm, to assign each node (i.e.~seed cluster, and by extension its constituent data points) to a final set of nested clusters. The Louvain method outputs a hierarchy of nested clusters at different resolutions, the lowest (i.e.~coarsest) resolution of which locally maximizes modularity \cite{Newman2004modularity}. Although I used the Louvain method for demonstration, other algorithms, such as Infomap \cite{Rosvall2008infomap2,Rosvall2011infomapM} or spectral clustering \cite{vonluxburg2007_spectral-clust-tutorial}, could also be applied to the ASTRICS similarity matrix/graph.



\section{Results}

\subsection{ASTRICS facilitates visualization and clustering of CyTOF data}

To demonstrate the potential for ASTRICS to facilitate and improve analysis of HD FC or CyTOF data in molecular biology, I applied a seed--ASTRICS--clu/vis pipeline (with ``seed'' = either $K$-means or the SOM step from FlowSOM, ``clu'' = the Louvain method, and ``vis'' = force-driected layout, all as described above) to Sample 01 from the publicly available CyTOF data published by Samusik et al.~\cite{Samusik2016}. Henceforth, this Sample 01 data is referred to as ``Samusik01''. Weber and Robinson also used this exact data in their comparison of clustering algorithms for FC and CyTOF data \cite{Weber2016}. In their study, FlowSOM was the best performing clustering algorithm for this data.

Samusik01 consists of measurements of the expression levels of 39 proteins on the surfaces of $86,864$ bone marrow cells from a C57BL/6J mouse (i.e.~$86,864$ observations with 39 dimensions or variables). The cells had been manually assigned to 24 known, immunologically distinct populations (clusters) by the original authors via manual gating according to known characteristics of the 24 cell types. 
Manual gating is an expert-driven process of scatter plotting single-cell cytometry data for two proteins (variables) at a time and drawing boundaries (``gates'') around cells (data points) to group them by cell type based on prior knowledge. It is the standard practice for classifying cells from FC or CyTOF data. 
Hence, the 24 manually gated populations can be regarded as ground-truth cluster labels against which computationally derived clusters can be compared in order to test computational methods for clustering. Of the $86,864$ cells, $53,173$ (61\%) were classified by manual gating; the remaining $33,691$ (39\%) were left unassigned. Nevertheless, I applied the seed--ASTRICS--clu/vis pipeline to all $86,864$ cells in the full 39-dimensional space.

Before applying the pipeline, I preprocessed the data by applying the arsinh transform with a cofactor of 5 (i.e.~$X \mapsto \asinh(X/5)$), as is standard for CyTOF data \cite{Nowicka2019_cytof-workflow}. This transformation makes the data more akin to a normal distribution in each dimension. I then normalized the data in each dimension by the width of the smallest interval containing 50\% of the positive-valued data, a robust estimate of scale \cite{Rousseeuw1993_alternatives-to-mad}. For the seed clustering step, I used $K=144$ (prior to any merging of seed clusters by my variant of centroid linkage) because 144 is a square number with nice factorization properties, which allowed different rectangular grid shapes to be used for the SOM without changing the total number of SOM nodes (for the SOM from FlowSOM, the height and width of a rectangular grid must be specified rather than $K$, so $K = \mathrm{height} \times \mathrm{width}$). I tested both the $\ell^2$ and $\ell^1$ norms as the distance measure used for seed clustering, but the figures show results using the $\ell^2$ norm, which is the default for both $K$-means and FlowSOM. I also tested the correlation distance measure in $K$-means seed clustering, but it consistently yielded inferior seed cluster purity and overall results for Samusik01 (results not shown).

For $K$-means seed clustering, I performed 10 clustering repeats using random initialization by the $K$-means++ algorithm \cite{Arthur2007_kmeans++} and then selected the clustering with the smallest value of its objective function (e.g.~the sum of squared Euclidean distances from each point to its cluster centroid if using the $\ell^2$ norm as the distance measure). For SOM seed clustering, I repeated the SOM step of FlowSOM six times using grids of size $24\times6$, $16\times9$, $12\times12$ (twice), $9\times12$, and $6\times24$. I then computed the mutual information for all possible pairs of SOMs and the Shannon entropy of each individual SOM (note that the entropy of a SOM is equal to its mutual information with itself). I subsequently selected as the seed clustering the SOM with the greatest sum of entropy and mutual information over all relevant pairs of SOMs. I did this as a fast, rudimentary form of ensemble clustering. Each SOM generated by FlowSOM is usually different (like the independent repeat runs of $K$-means), and any individual SOM could be anomalous. However, the SOM with the highest entropy and shared information content among all generated SOMs is conceptually like the median SOM and is therefore unlikely to be anomalous.

\begin{figure}[htbp]
    \centering
    \vspace{-0.3cm}
    \includegraphics[width=\textwidth]{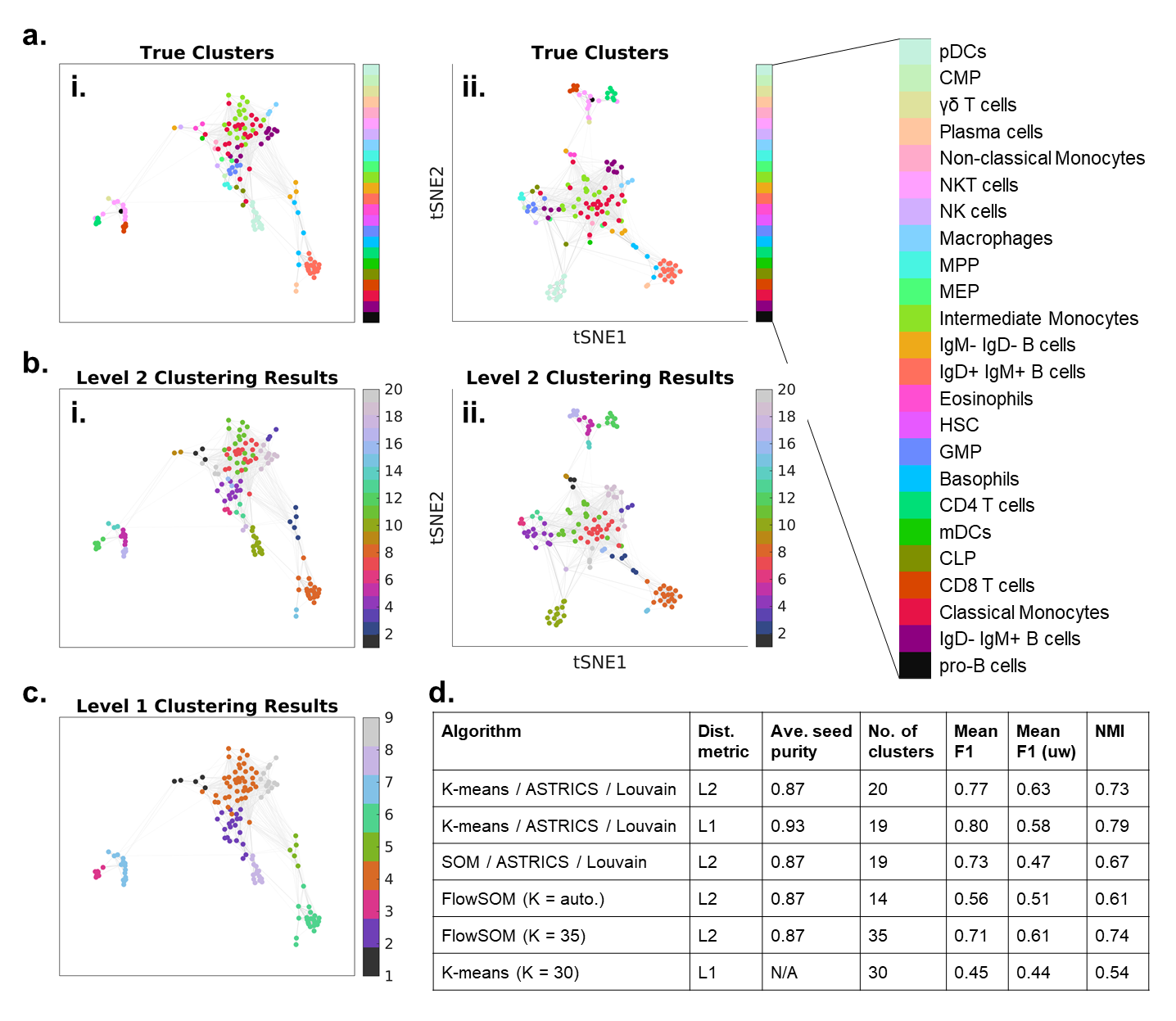}
    \vspace{-0.8cm}
    \caption[Visualization and clustering results using ASTRICS for 39-dimensional Samusik01 CyTOF data.]{\textbf{Visualization and clustering results using ASTRICS for 39-dimensional Samusik01 CyTOF data.}
	\textbf{(a)} ASTRICS similarity graph for `seed clusters' of cells (one seed cluster = one node in graph) visualized using force-directed layout (i) or Barnes-Hut t-SNE map of seed cluster centroids using \textit{perplexity} = 20 (ii). Dataset contained $86,864$ cells, each with CyTOF measurements of the expression levels of 39 proteins on the cell surfaces. Seed clusters were determined by $K$-means using $K=144$ and the Euclidean (L2) distance measure. Nodes are coloured according to the most common cell type within the seed clusters they represent.
	Cell type abbreviations: pDCs = plasmacytoid dendritic cells, CMP = common myeloid progenitor, NK = natural killer, MPP = multipotent progenitors, MEP = megakaryocyte–erythroid progenitor, HSC = hematopoietic stem cells, GMP = granulocyte-monocyte progenitor, mDCs = myeloid dendritic cells, CLP = common lymphoid progenitor.
	\textbf{(b--c)} Two resolutions of clusters found by applying the Louvain method to the ASTRICS similarity graph. Colours distinguish clusters.
	\textbf{(d)} Assessment of results from different clustering methods by mean F-measure (uw = unweighted) and normalized mutual information (NMI) compared to manually gated populations. 1~=~100\% match; 0~=~0\% match. Where applicable, the table relates to the finest resolution of clusters output by the Louvain method, which was Level 2, shown in (b), for the clustering in the figure. The second column states the distance metric used in clustering: $\mbox{L2} = \ell^2$ (Euclidean distance) or $\mbox{L1} = \ell^1$.}
    \label{fig:ASTRICS_Samusik01CyTOF}
\end{figure}

To qualitatively assess the visualization generated by force-directed layout of the ASTRICS similarity graph, I plotted it with nodes coloured according to their ground-truth assignments to one of the 24 cell types (\hlcy{Figure}~\ref{fig:ASTRICS_Samusik01CyTOF}a(i)). The seed clusters represented by the nodes in the figures were obtained using $K$-means, and their ground-truth assignments were determined by simple majority voting of the manually gated cells (the cells left unassigned by manual gating did not contribute to the voting). Shown alongside the force-directed layout for comparison is a t-SNE map of the seed cluster centroids generated using the Barnes-Hut t-SNE algorithm \cite{vandermaaten2014_bh-tsne} with the perplexity parameter set equal to 20 (\hlcy{Figure}~\ref{fig:ASTRICS_Samusik01CyTOF}a(ii)). 
We can see that force-directed layout of the ASTRICS similarity graph and t-SNE applied to the seed cluster centroids produced qualitatively similar layouts of the nodes. Both layouts have pro-B cells and all T cells (CD4$^+$ T cells, CD8$^+$ T cells, NKT cells, and \gam\delt\ T cells) grouped closely together in one arm and another arm with IgD$^+$ IgM$^+$ B cells, basophils, plasma cells, and IgD$^-$ IgM$^-$ B cells. Also in both layouts, plasmacytoid dendritic cells (pDCs) form another distinguishable population while the remaining cell types are grouped together in a larger central cloud of nodes. Since t-SNE is a leading method to visualize CyTOF data, we can conclude that force-directed visualization based on ASTRICS is a very effective method to visualize CyTOF data at the resolution of seed clusters. Moreover, except for choosing the resolution for the seed clusters, my method has the advantage of not requiring the user to tune any parameters such as the perplexity in t-SNE.

In the shown figures, the three monocyte populations, especially the classical and intermediate monocytes, are largely mixed together and difficult to distinguish from each other. However, they were almost completely distinguishable from each other in both the force-directed and t-SNE layouts when I used the $\ell^1$ instead of $\ell^2$ norm as the distance measure in the seed clustering by $K$-means (not shown).
In fact, this resulted in better separation of distinct cell types in general in both force-directed and t-SNE visualizations, likely due to higher purity seed clusters (see \hlcy{Figure}~\ref{fig:ASTRICS_Samusik01CyTOF}d). Unfortunately, it also resulted in the rare pro-B cell population being completely lost among other populations in the seed clustering (i.e.~pro-B cells did not form the most abundant manually gated cell type within any seed cluster). That said, I note that the single pro-B cell node resulting from the use of $\ell^2$ in the $K$-means seed clustering already had the second lowest purity (33\%) of all seed clusters. Furthermore, I found that the $K$-means seed clustering was less likely to converge within 200 iterations when using $\ell^1$ instead of $\ell^2$. The results suggest that, for $K$-means, using $\ell^1$ instead of $\ell^2$ produces higher quality seed clusters overall (even if it does not always converge), but possibly at the cost of losing rare populations. Consideration should therefore be given to the distance measure used during seed clustering in a seed--ASTRICS--clu/vis pipeline. I suggest that, before applying such a pipeline to new data, potential algorithms and distance measures for the seed clustering first be tested on similar data with known classifications of observations, as I have done here.

The visualizations presented here effectively validate ASTRICS as a suitable method to measure similarity between clusters of cells in CyTOF data. This validation arises from not only the qualitative similarity between the ASTRICS-based visualization and t-SNE but also the agreement of the visualization with rational expectations: biologically similar cell types are visually clustered together. For instance, the T cell subsets are all close together in the visualization, as are the monocyte subsets. Nonetheless, I also tested the suitability of the ASTRICS similarity matrix as input to a clustering algorithm as per the description of the general seed--ASTRICS--clu/vis pipeline using the Louvain method as an example.

Applying the Louvain method to the ASTRICS similarity graph shown in \hlcy{Figure}~\ref{fig:ASTRICS_Samusik01CyTOF}a yielded nested clusters over two resolutions (\hlcy{Figure}~\ref{fig:ASTRICS_Samusik01CyTOF}b--c). The seed clusters (the nodes of the graph) themselves form another resolution. I quantified the quality of the finest resolution of clusters returned by the Louvain method compared to the ground-truth clusters (i.e.~the manually gated cell populations) using the mean F-measure \cite{Aghaeepour2013meanfmeasure} and normalized mutual information (NMI) \cite{Danon2005nmi}. These quantities range from 0 to 1 with a value of 1 indicating perfect agreement between the algorithmically derived and ground-truth cluster assignments of observations. I computed both the usual mean F-measure, in which the contribution of each ground-truth cluster is weighted proportionally to its size, and an unweighted version to which all ground-truth clusters contribute equally regardless of their size. I computed the unweighted version to give rare cell populations equal importance to large cell populations, similar to the assessment approach taken by Weber and Robinson \cite{Weber2016}. All quantities were computed using only the manually gated cells; cells not assigned to any manually gated population were completely discounted from all computations of mean F-measure, NMI, and purity. This clustering assessment was performed for different algorithms. I tested all combinations of $\ell^2$ or $\ell^1$ as the distance measure with $K$-means or the SOM step from FlowSOM as the algorithm for seed clustering. For comparison, I also tested clustering using just FlowSOM, with both automatic and user selection of $K$, or just $K$-means. Selected results of this analysis are summarized in \hlcy{Figure}~\ref{fig:ASTRICS_Samusik01CyTOF}d.
In a similar vein to the visualizations, I found that the overall best clustering results, as assessed by NMI and the regular mean F-measure, were generated by the $K$-means--ASTRICS--Louvain pipeline using the $\ell^1$ distance measure in $K$-means. This particular pipeline returned 19 clusters. The same pipeline using $\ell^2$ had lower NMI and regular mean F-measure scores, but it returned one additional cluster and a higher unweighted mean F-measure. Supporting my earlier observations, this suggests that $\ell^2$ is the better choice of distance measure for $K$-means seed clustering if finding rare cell populations is important, though $\ell^1$ is better for most clusters. 

Nonetheless, for both the $\ell^1$ and $\ell^2$ distance measures, $K$-means--ASTRICS--Louvain yielded much better clustering results than using just FlowSOM with automated selection of the final number of clusters. By manually tuning the final number of clusters in FlowSOM, I was able to obtain results comparable to those generated by $K$-means--ASTRICS--Louvain, but $K$-means--ASTRICS--Louvain was generally still better overall. Surprisingly, the use of the SOM from FlowSOM as the seed clustering step in the seed--ASTRICS--Louvain pipeline produced worse final clustering results than using $K$-means for seed clustering, despite the seed clusters being of similar purity in both cases. I point out, notwithstanding, that this was still slightly better overall than FlowSOM with automated selection of the number of clusters. Recall that FlowSOM was previously found by Weber and Robinson \cite{Weber2016} to be the best available clustering algorithm for Samusik01. Unquestionably, using just $K$-means on its own for clustering produced worse results than seed--ASTRICS--Louvain and also worse results than FlowSOM. The results of the assessment here clearly show that ASTRICS is capable of not only facilitating but also improving clustering analysis of CyTOF data.

\subsection{Application to MNIST digital images demonstrates versatility of ASTRICS}

\begin{figure}[htbp]
    \centering
    \includegraphics[width=0.67\textwidth]{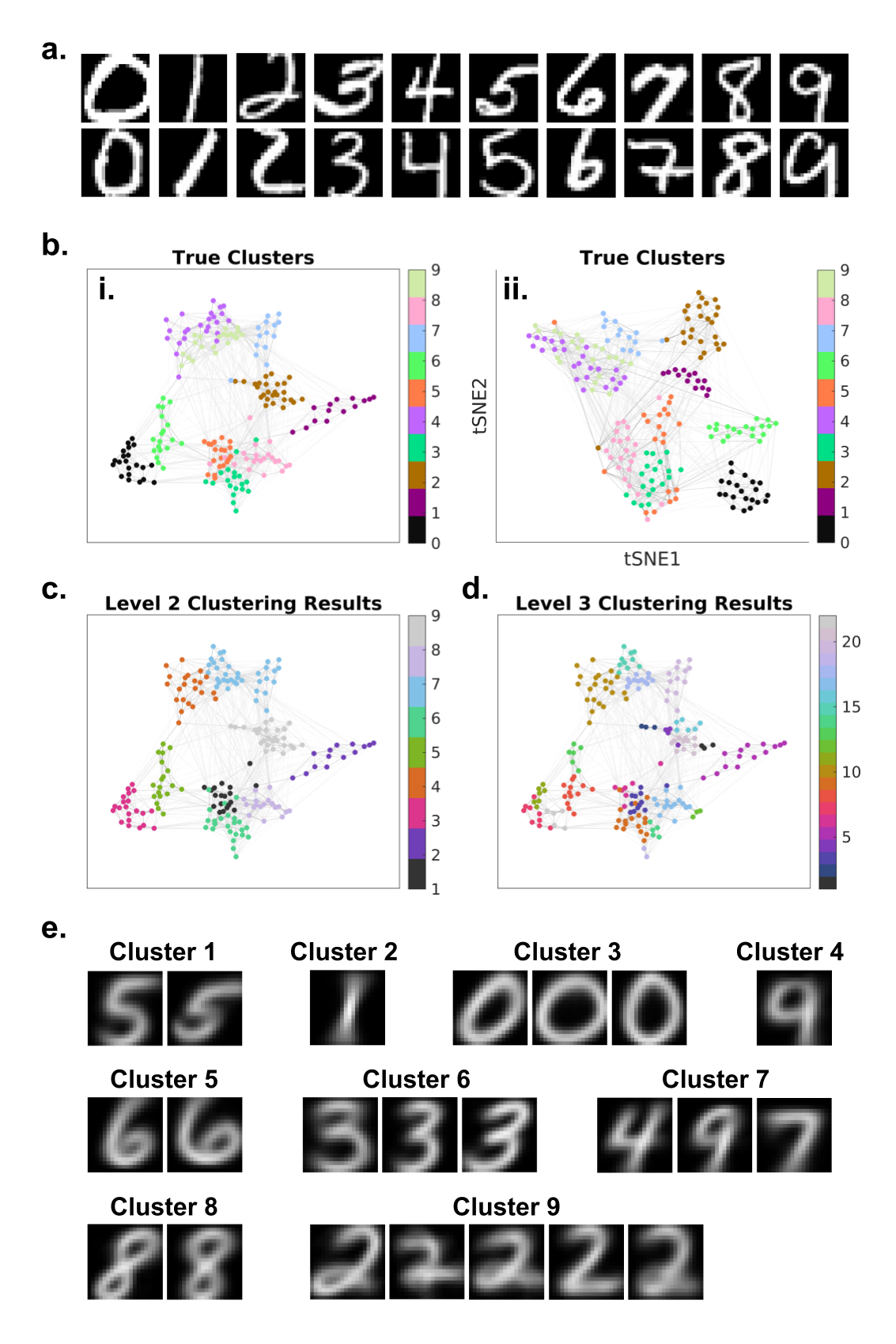}
    \caption[Visualization and clustering results using ASTRICS for 400-dimensional MNIST images of handwritten digits.]{\textbf{Visualization and clustering results using ASTRICS for 400-dimensional MNIST images of handwritten digits.} \textbf{(a)} Example images from the MNIST dataset. \textbf{(b)} Visualization by force-directed layout of the ASTRICS similarity graph (i) or Barnes-Hut t-SNE (\textit{perplexity} = 20) of seed cluster centroids (ii) for $K=200$ seed clusters of images from a set of $20,000$ MNIST images of handwritten digits. Seed clusters were determined by $K$-means using the correlation distance measure. Each node is coloured according to the most common ground-truth digit (0 to 9) within the seed cluster it represents. \textbf{(c--d)} The second (c) and third (d) coarsest resolutions of clusters found by applying the Louvain method to the ASTRICS similarity graph in panel (b). Colours distinguish distinct clusters. Cluster numbers have no correspondence to the true digits because the clustering is unsupervised. \textbf{(e)} Means of the 22 clusters in panel (d) grouped according to the clusters in panel (c). Cluster numbers correspond to the cluster numbers in panel (c).}
    \label{fig:ASTRICS_MNIST}
\end{figure}

To demonstrate that ASRTICS is suitable for a broad variety of data, I also applied the $K$-means--ASTRICS--Louvain/force pipeline, where ``force'' refers to the use of force-directed layout for visualization, to digital images from the MNIST database \cite{Lecun1998_mnist}. This is a collection of grayscale digital images, 20 pixels $\times$ 20 pixels in size, of handwritten single-digit numbers (\hlcy{Figure}~\ref{fig:ASTRICS_MNIST}a). The 10 unique numbers 0--9 were regarded as being ground-truth cluster labels. Again, I compared the visualization produced by $K$-means--ASTRICS--force to a Barnes-Hut t-SNE map of the seed cluster centroids, and I quantified the quality of clustering results from $K$-means--ASTRICS--Louvain compared to ground truth using the mean F-measure and NMI.

To apply the methods to the MNIST images, the images were reshaped into 400-dimensional row vectors. They were not preprocessed in any other way. 
For seed clustering, I used $K=200$. I tested the $\ell^1$, $\ell^2$, and correlation distance measures on a development dataset of $20,000$ images and found that the correlation distance measure produced the best results for the MNIST data. This contrasts my observations for the CyTOF data and suggests that the best choice of distance measure will depend on the dataset. I therefore reiterate my earlier recommendation to test different distance measures on similar data with known ground-truth cluster labels before applying clustering methods to new data.

After determining the best distance measure, I applied $K$-means--ASTRICS--Louvain/force to a completely different set of $20,000$ MNIST images in order to evaluate the pipeline. \hlcy{Figure}~\ref{fig:ASTRICS_MNIST}b shows (i) a force-directed layout of the resulting ASTRICS similarity graph alongside (ii) a Barnes-Hut t-SNE map of the seed cluster centroids. Just as for the CyTOF data, we see substantial qualitative similarity between the two layouts. Many of the ground-truth clusters representing distinct digits are visually distinct from each other. Even the digits that are less discernible from each other in the layouts cluster together logically based on typographical similarities: the numbers ``3'', ``5'' and ``8'' cluster together, as do ``4'' and ``9''. Typographical similarities also explain the closeness of the digits ``0'' and ``6'', and the digits ``1'' and ``2'' in both layouts. Furthermore, force-directed layout of the ASTRICS similarity graph conveys a sense of global relationships between digits: the two most distant digits in the force-directed layout are ``0'' and ``1'', the two most typographically dissimilar digits. This is not apparent in the t-SNE map, which is not surprising because t-SNE preserves only local relationships and cannot be used to infer global relationships. This illustrates an advantage of $K$-means--ASTRICS--force over t-SNE when visualizing HD data at the resolution of fine-grained clusters instead of the original data points is acceptable. The effectiveness of the ASTRICS-based visualization approach for the MNIST data in addition to CyTOF data demonstrates that ASTRICS is a practical measure of inter-cluster similarity not limited to one type of data.

The Louvain method clustering step of the $K$-means--ASTRICS--Louvain pipeline generated a hierarchy of clusters over three resolutions, in addition to the initial seed cluster resolution. The middle resolution (\hlcy{Figure}~\ref{fig:ASTRICS_MNIST}c) consisted of 9 clusters, the closest to the true number of distinct digits (i.e.~10). It achieved a mean F-measure of 0.78 and an NMI of 0.74. For comparison, applying just $K$-means using the correlation distance measure and the correct number of clusters $K=10$ to the same $20,000$ images achieved a mean F-measure of 0.57 and an NMI of 0.50. The lowest resolution of clusters returned by the Louvain method had 8 clusters: the clusters most closely matching the digits ``5'' and ``9'' in the 9-cluster resolution (clusters 1 and 8 in \hlcy{Figure}~\ref{fig:ASTRICS_MNIST}c) were merged together, but otherwise the two lowest-resolution clusterings were the same. It is noteworthy that almost the correct number of clusters could be obtained without having to specify the final number of clusters using $K$-means--ASTRICS--Louvain, and that this approach yielded far superior results to performing just $K$-means with the true number of clusters. Moreover, the comparatively poor performance of $K$-means on its own cannot be attributed to variability in the number of images of each digit in the dataset because all 10 digits were present in very similar proportions. Thus, the dimensionality is likely to be a key factor. These results clearly show that a seed--ASTRICS--clu pipeline can expand the utility of generic clustering algorithms to higher dimensions by using ASTRICS to mitigate the curse of dimensionality between a fine-grained initial clustering step and a graph-based final clustering step.

At the finest resolution output by the Louvain method, there were 22 clusters (\hlcy{Figure}~\ref{fig:ASTRICS_MNIST}d). To visually assess the cluster hierarchy, I computed the mean image for each of these 22 clusters (i.e.~their centroids) and grouped them according to the next level of clustering (\hlcy{Figure}~\ref{fig:ASTRICS_MNIST}e). In general, the 22 clusters corresponded to different writing styles, especially italic and upright, for the distinct digits in the MNIST dataset. There were three notable exceptions in which images of different digits were clustered together. First, the sole subcluster of Cluster 4 in \hlcy{Figure}~\ref{fig:ASTRICS_MNIST}e contained images of both ``4'' and ``9''. However, it is visually apparent that these were upright writing styles. The italic images of ``4'' and ``9'' formed separate subclusters of Cluster 7. Second, the first subcluster of Cluster 6 in \hlcy{Figure}~\ref{fig:ASTRICS_MNIST}e contained images of ``3'', ``5'', and ``8''. 
Both of these exceptions can be explained by obvious visual similarities between some written styles of the distinct digits. 
Last, none of the 22 clusters appear to represent ``7'' written with a horizontal crossbar, as in the bottom row of \hlcy{Figure}~\ref{fig:ASTRICS_MNIST}a. Only one of the 200 seed clusters strongly represented this style of writing ``7'', but the Louvain method is very unlikely to return any single-node clusters except for zero-degree nodes. This `crossbar-7' seed cluster had nonzero similarity with seed clusters of other digits and subsequently clustered with images of ``2'' in the second subcluster of Cluster 9 in \hlcy{Figure}~\ref{fig:ASTRICS_MNIST}e.
It can be deduced from these observations that variability in writing styles for specific digits is significant compared to the differences between distinct digits. Indeed, some of the handwritten digits among the MNIST images are ambiguous even to the human eye.

At the next, coarser level of clusters, the 22 finer-resolution clusters mostly clustered together by actual digit. For example, italic and upright styles of ``5'' and ``8'' formed separate fine-resolution clusters, which in turn formed a ``5'' cluster and an ``8'' cluster (Clusters 1 and 8 respectively in \hlcy{Figure}~\ref{fig:ASTRICS_MNIST}e). The obvious exception to this is Cluster 7 in \hlcy{Figure}~\ref{fig:ASTRICS_MNIST}e, in which subclusters whose means are recognizably the digits ``4'', ``9'' and ``7'' were clustered together. However, despite being recognizable as three distinct digits, it is easy to see why these three clusters were clustered together at coarser cluster resolutions: their constituent images share a long slanted line at the rightmost side of the digit. By applying my pipeline combining $K$-means, ASTRICS, and the Louvain method to MNIST data and subsequently viewing the cluster centroids as in \hlcy{Figure}~\ref{fig:ASTRICS_MNIST}e, it was possible to visually explore variability and similarities between handwriting styles of numerals. Furthermore, apparent inaccuracies in the results could be rationally explained. From the strong and rational performance of the $K$-means--ASTRICS--Louvain/force clustering and visualization pipeline on MNIST image data in addition to CyTOF data, I conclude that ASTRICS is a versatile measure of inter-cluster similarity, suitable for multiple types of HD data and a range of dimensionalities.

\subsection{Global feature extraction is still helpful in very high dimensions}

To test whether ASTRICS is suitable for very HD data with thousands of dimensions, I next applied a seed--ASTRICS--clu/vis pipeline to text documents from the \textit{20 Newsgroups} dataset \cite{20newsgroups_rainbow, 20newsgroups_home}, a corpus of $18,846$ newsgroup documents spread roughly evenly across 20 different topics. The documents are separated into a training set of $11,314$ documents and a test set of $7,532$ documents. The 20 topics constitute ground-truth cluster labels for the documents. All headers were removed from the documents because these included the exact ground-truth topic for each document.
In order to apply my methods, or any clustering algorithm, to text documents, the documents had to first be converted to numerical vectors, a process known as \textit{vectorization}. This was achieved using a `bag of words' approach, which encodes the frequency of different `words' in the documents. Here, `word' means any sequence of characters (i.e.~string) without spaces or breaks. Subsequently, I applied a seed--ASTRICS--clu/vis pipeline to the vectorized documents.

Using only the training subset of the \textit{20 Newsgroups} data at first, I trialled various combinations of parameters and options for the vectorization of documents and for my pipeline. I eventually settled on the following approach. 
First, I \textit{tokenized} documents (i.e.~converted them to lists of \textit{tokens}, or strings) in Python by applying the \texttt{word\_tokenize} function from Natural Language Toolkit (NLTK) \cite{nltk}, removing words shorter than three characters, and \textit{stemming} words using the Snowball stemmer from NLTK. Stemming reduces words to their stems so that, for example, different tenses of a verb, or singular and plural forms of a noun, reduce to the same token. I also removed ubiquitous words such as ``and'' and ``the'', called \textit{stop words}, which are generally considered to be uninformative for clustering and machine learning. Finally, to convert the tokenized documents into numerical vectors, I used the term frequency--inverse document frequency (TF-IDF) vectorizer (\texttt{TfidfVectorizer}) from scikit-learn \cite{scikit-learn}, removing any tokens that appeared in fewer than three documents during the process. For each tokenized document, \texttt{TfidfVectorizer} creates a row vector in which each column records a weighted frequency of occurrence of a particular token. The ``TF'' part normalizes token counts by document length (number of tokens). The ``IDF'' part inversely weights tokens by the logarithm of the fraction of documents in which they appear, which reduces the importance of words that appear in many documents. The dimensionality of the vectorized text data is the total number of tokens recorded in the corpus. Note that this is not fixed because it depends on the documents in the corpus. Using the vectorization approach described, the training subset, test subset, and complete set of documents had $29,737$ dimensions, $21,683$ dimensions, and $42,823$ dimensions respectively.

Seed clustering was performed using $K$-means (this time using 20 repeats) with $K=200$ and the $\ell^1$ distance measure followed by my variant of centroid-linkage with a minimum cluster size of 8 data points. Prior to running $K$-means, 50 features were extracted by SVD, often called latent semantic analysis in the context of text data, to reduce the dimensionality globally. I then either applied ASTRICS to the seed clusters back in the original very HD space or performed global feature extraction by SVD before applying ASTRICS. As usual, I used force-directed layout to visualize the ASTRICS similarity graph and the Louvain method for the final clustering step.

For the training set of vectorized text documents, reducing the dimensionality of the data to 20, 50, 100, 150, 200, 500, or $1,000$ features by SVD before applying ASTRICS consistently yielded better clustering results, in terms of the mean F-measure and NMI, than applying ASTRICS in the full $29,737$ dimensions. The actual choice of the reduced number of features had relatively little impact on the final results measured by the mean F-measure or NMI. Performing feature extraction before ASTRICS resulted in mean F-measure scores between 0.44 and 0.50 and NMI scores mostly between 0.46 and 0.48 for the level of clusters from the Louvain hierarchy that achieved the closest match to the ground-truth clusters. Three replicates of clustering without the feature-extraction step before ASTRICS resulted in mean F-measure scores from 0.38 to 0.43 and NMI scores from 0.43 to 0.44. Notably, the seed clusters themselves had an NMI value of 0.45 in all three replicates. 
It should also be noted that the seed clusters had a weighted mean purity between 60\% and 61\% and an unweighted mean purity between 66\% and 68\%, which indicates that there was already considerable mixing of different newsgroups within the seed clusters. 
Based on the results for the training dataset and the increasing computational cost of performing SVD as the desired number of features increases, 100 was chosen as the number of features to proceed with for the test dataset.

\hlcy{Figure}~\ref{fig:ASTRICS_20Newsgroups}a shows a force-directed layout and Louvain clustering results for 
seed clusters in the test set when ASTRICS was applied after reduction of the dimensionality to 100 features. The finest resolution (level 3) of the Louvain clustering scored 0.45 for the mean F-measure and 0.47 for the NMI. Two additional Louvain clustering replicates (not shown) scored 0.45 and 0.43 for the mean F-measure and 0.47 and 0.46 for the NMI.
\hlcy{Figure}~\ref{fig:ASTRICS_20Newsgroups}b shows analogous results when ASTRICS was instead applied to the same seed clusters in the full $21,683$ dimensions. In this case, the finest resolution (level 2) of the Louvain clustering scored 0.43 for the mean F-measure and 0.44 for the NMI. Two additional Louvain clustering replicates each scored 0.43 for the mean F-measure and 0.45 for the NMI.
The seed clusters had a weighted mean purity of 61\% and an unweighted mean purity of 66\%. For reference, their mean F-measure was 0.26 and their NMI score was 0.46.
Where Louvain clusters in either graph overlapped more than one of the 20 newsgroup topics, which was detrimental to the NMI and mean F-measure scores, nodes that clustered together were often rationally related, such as nodes representing different topics about computers.
Furthermore, the force-directed layouts of the two graphs were qualitatively similar in that both had a star-like structure with crowding of nodes in the centre and nodes representing the same or similar newsgroup topics generally organized close together. Nevertheless, the layout had clearer visual organization of nodes into clusters along `arms' of the graph when ASTRICS was preceded by global feature extraction (\hlcy{Figure}~\ref{fig:ASTRICS_20Newsgroups}a) than when all $21,683$ dimensions were used (\hlcy{Figure}~\ref{fig:ASTRICS_20Newsgroups}b).
The graph also had fewer zero-degree nodes when ASTRICS was preceded by global feature extraction (18 zero-degree nodes in \hlcy{Figure}~\ref{fig:ASTRICS_20Newsgroups}a versus 39 in \hlcy{Figure}~\ref{fig:ASTRICS_20Newsgroups}b). 
This shows that ASTRICS more often discovered feature subspaces that discriminated between seed clusters of documents discussing the same topic when it was allowed to use all of the data dimensions. This perhaps reflected different writing styles or lexicons between authors of documents about the same topic. 
However, one could argue that ASTRICS overfitted to noise if the goal was to cluster documents more broadly by topic. Thus, global feature extraction helped ASTRICS by removing some of the irrelevant intra-topic noise in the vectorized text data that would otherwise be detected by local DR.

\begin{figure}[htbp]
    \centering
    \includegraphics[width=0.94\textwidth]{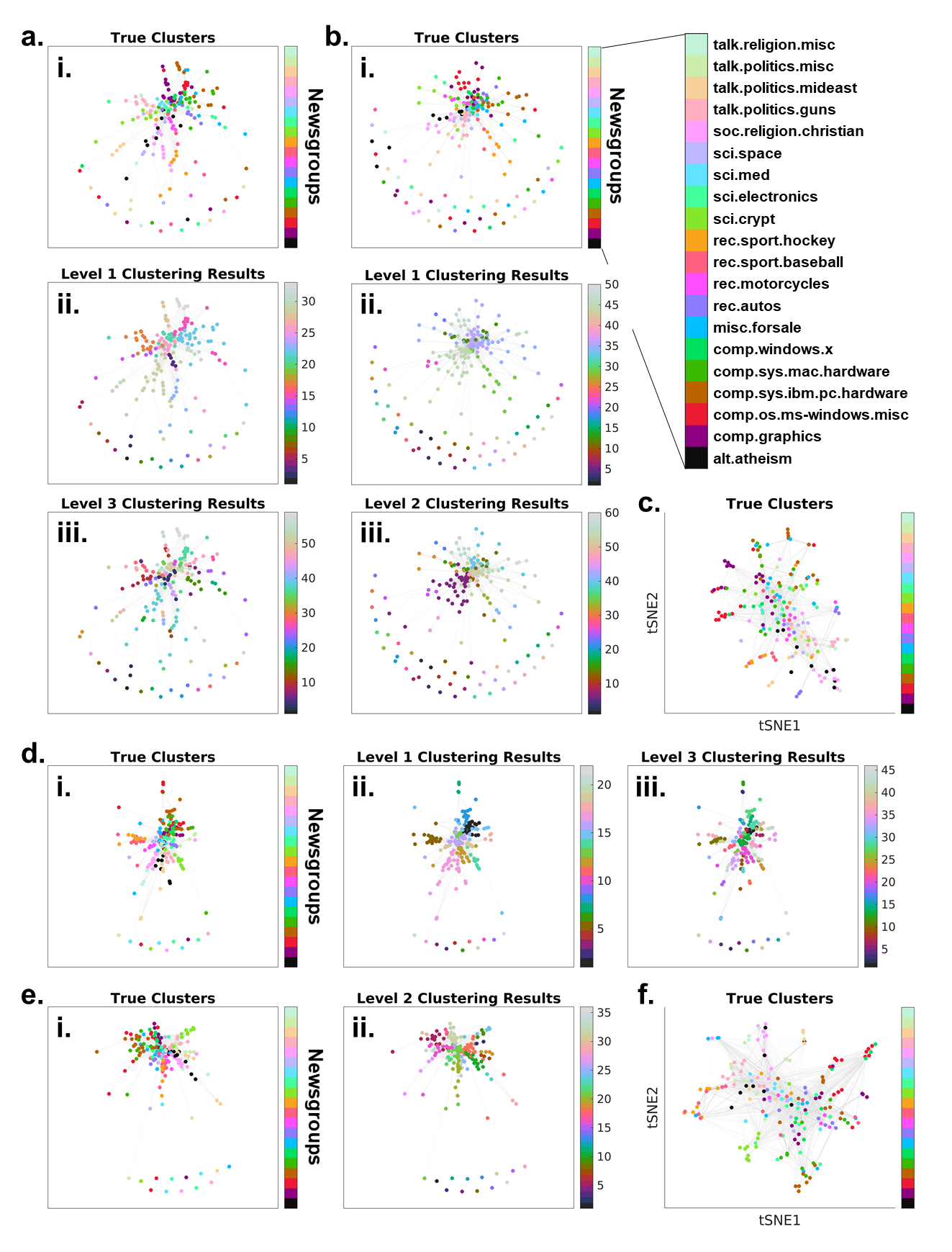}
    \caption[Visualization and clustering results using ASTRICS with and without global feature extraction for very HD newsgroups text documents.]{\textbf{Visualization and clustering results using ASTRICS with and without global feature extraction for very HD newsgroups text documents.} 
    Caption continues~\dots}
    \label{fig:ASTRICS_20Newsgroups}
\end{figure}

\begin{figure}[!tbp]
    \centering
    \contcaption{For the $21,683$-dimensional test set (a--c) and $42,823$-dimensional complete set (d--f) of vectorized text documents from the \textit{20 Newsgroups} dataset, seed clusters were defined by performing $K$-means clustering using the $\ell^1$ distance measure and $K=200$ in a 50-dimensional feature space. If any $K$-means clusters contained fewer than 8 data points, the smallest cluster was merged to its nearest neighbouring cluster based on distances between centroids in the 50-dimensional feature space and this process was repeated until all clusters contained at least 8 data points. This resulted in 191 seed clusters for the test set and 199 for the complete set. Each node in any panel of this figure represents one seed cluster. 
    \textbf{(a)} Force-directed layout of the graph generated by applying ASTRICS to the seed clusters in the test set in a 100-dimensional feature space obtained by global SVD. Nodes are coloured to show the most common ground-truth topic of the documents within each seed cluster (i) and the coarsest (ii) and finest (iii) levels of clusters obtained by applying the Louvain method to the graph.
    \textbf{(b)} Same as (a) except that ASTRICS was performed in the full $21,683$-dimensional space for the test set. Note the larger number of zero-degree nodes in (b) than in (a).
    \textbf{(c)} Barnes-Hut t-SNE map of the centroids of the seed clusters in the test dataset using \textit{perplexity} = 20. Nodes are coloured as in a(i) and b(i). The edges are the same as the edges in the graph in (a).
    \textbf{(d--f)} Repeat of (a--c) respectively for the $42,823$-dimensional complete set of vectorized newsgroups documents (training and test sets combined), except that the finest level of Louvain clusters is omitted from panel (e).}
\end{figure}

A Barnes-Hut t-SNE map of the seed cluster centroids, which involved an initial linear reduction to 20 dimensions to prevent overfitting, did not exhibit central crowding of nodes and therefore made it easier to see more of the individual nodes (\hlcy{Figure}~\ref{fig:ASTRICS_20Newsgroups}c). Indeed, crowding is a common problem of DR methods that t-SNE explicitly addresses \cite{vandermaaten2008_tsne}. 
However, it made some of the true clusters appear more fragmented or scattered. For example, five nodes representing primarily documents from the ``rec.autos'' newsgroup topic were clustered together in both the force-directed layout and Louvain clustering of the ASTRICS similarity graph but were separated into two distant regions of the t-SNE map.
On the other hand, nodes that had degree zero and therefore formed singletons in the ASTRICS similarity graph were often located near nodes representing the same or similar topics in the t-SNE map. 
Hence, force-directed layout of the ASTRICS similarity graph for the seed clusters and t-SNE applied to the centroids of the seed clusters yielded complementary visualizations. Overlaying the edges of the ASTRICS similarity graph on the t-SNE map (\hlcy{Figure}~\ref{fig:ASTRICS_20Newsgroups}c) added information about the inter-relatedness of nodes that were distant or scattered in the t-SNE map.

Unfortunately, however, the $K$-means--ASTRICS--Louvain clustering pipeline was not able to improve on the clustering results that could be achieved using just $K$-means for the test subset of the vectorized \textit{20 Newsgroups} data. Clustering by $K$-means using $K=20$ and the $\ell^1$ distance measure in a 50-dimensional feature subspace obtained by SVD achieved a mean F-measure of 0.49 and an NMI of 0.45. 
Nevertheless, it is noteworthy that $K$-means--ASTRICS--Louvain achieved comparable clustering results to $K$-means without having to specify the final number of clusters. Also, the $K$-means--ASTRICS--Louvain/force approach has the advantage of providing a visualization alongside the clustering results.

To investigate whether the seed clustering was the limiting step in the $K$-means--ASTRICS--Louvain pipeline, I used the ground-truth topic label for each document to generate seed clusters all of 100\% purity. Similarities between pure seed clusters were then computed as usual using ASTRICS, either with or without prior global reduction of the dimensionality to 100 features by SVD, and clusters were obtained from the resulting ASTRICS similarity graph using the Louvain method. The mean F-measure and the NMI were calculated for the lowest resolution of Louvain clusters. This test was performed using two different methods to generate pure seed clusters. 

First, the vectorized text documents within each topic category in the \textit{20 Newsgroups} test set were randomly assigned to 10 clusters per topic. When ASTRICS was applied to these seed clusters in the full-dimensional space, three replicates of Louvain clustering (each of which returned the best of 100 runs) yielded identical results for two of the replicates, scoring 0.70 for the mean F-measure and 0.87 for the NMI, and the third replicate scored 0.60 for the mean F-measure and 0.81 for the NMI. Either seven or six of the 20 topics were recovered with 100\% accuracy and others clustered together rationally: political topics clustered together, ``religion'' and ``atheism'' clustered together, and topics related to computers or electronics clustered together. The graph had a single connected component and there were many edges between very different topics.
When the dimensionality was reduced to 100 extracted features before applying ASTRICS, three replicates of Louvain clustering produced mean F-measure scores from 0.60 to 0.75 and NMI scores from 0.80 to 0.90. Seven to nine topics were recovered with 100\% accuracy and other topics again clustered together rationally.
This test using randomly generated seed clusters of 100\% purity validated ASTRICS as a useful similarity measure and showed that it is very effective for clusters that can overlap.

Second, pure seed clusters were generated by applying $K$-means clustering with $K=10$ in a 20-dimensional feature space to each of the 20 topics independently. The 20-dimensional feature space was also computed independently for each topic using SVD. Note that 28 of the 200 seed clusters defined this way contained fewer than 8 data points and the smallest one contained just three. Applying ASTRICS to these pure seed clusters in the full-dimensional space resulted in 121 zero-degree nodes (plus six additional clusters identified by the Louvain method), which suggests that the data were overfitted. When the dimensionality was globally reduced to 100 features by SVD before applying ASTRICS, the number of zero-degree nodes was reduced to 25, the mean F-measure scores for three replicates of Louvain clustering were between 0.42 and 0.44, and the NMI scores were between 0.59 and 0.60. Hence, global feature extraction prevented overfitting, but clustering results were not improved by increasing the purity of seed clusters defined by a hard-clustering algorithm, which typically results in non-overlapping clusters by construction.
Overall, the $K$-means--ASTRICS--Louvain results for artificially constructed seed clusters of 100\% purity show that the \textit{20 Newsgroups} data are noisy and have topics that overlap in an HD feature space. This reality, and not the purity of the seed clusters, was the limiting factor for obtaining good clustering results in terms of matching the known topic labels.

To test the $K$-means--ASTRICS--Louvain/force clustering and visualization approach on a larger, higher-dimensional data set, I repeated it for the $42,823$-dimensional set of all $18,846$ documents in the entire \textit{20 Newsgroups} dataset (even though this included the training set of documents on which I optimized my approach). The resulting visualizations and Louvain clusters with and without global extraction of 100 features by SVD before ASTRICS are shown in \hlcy{Figures}~\ref{fig:ASTRICS_20Newsgroups}d and~\ref{fig:ASTRICS_20Newsgroups}e respectively. A Barnes-Hut t-SNE map (which again involved initial linear DR to 20 dimensions) of the centroids of the same seed clusters is shown in \hlcy{Figure}~\ref{fig:ASTRICS_20Newsgroups}f for comparison. The seed clusters had weighted and unweighted mean purities of 58\% and 63\% respectively. 
The lowest resolution of Louvain clusters for the entire dataset was a marginally better match to the actual newsgroups when global DR to 100 dimensions was performed before ASTRICS (for each of three Louvain clustering replicates, $\mbox{mean F-measure} = 0.45$ and $\mbox{NMI} = 0.45$) than when ASTRICS was allowed to use all $42,823$ dimensions (for three Louvain clustering replicates: $\mbox{mean F-measure} = 0.41,\, 0.40,\, 0.41$; $\mbox{NMI} = 0.43,\, 0.42,\, 0.43$), though there was only a small difference in the number of zero-degree nodes this time (10 and 12 respectively). Also, the force-directed layouts of the ASTRICS similarity graphs were qualitatively similar to each other and to the layouts of the graphs for the test dataset.
The smaller numbers of zero-degree nodes in the graphs for the entire dataset compared to the graphs for just the test set were probably due to using the same value of $K$ for the seed clustering for both. This would have resulted in a greater average number of points per seed cluster in the larger complete dataset than the smaller test subset, which in turn would have reduced the extent of overfitting.
In conclusion, ASTRICS is a suitable measure of inter-cluster similarity for very HD data such as text documents, but it benefits from prior global feature extraction to prevent overfitting. Without global feature extraction, the local DR step of ASTRICS could identify subspaces that convincingly separate seed clusters, but those subspaces might be constructed from features that are ultimately not relevant for the problem being addressed. Hence, the power of global feature extraction should not be neglected even when utilizing local DR such as in ASTRICS.

\subsection{Comparison of local DR methods}

\begin{figure}[!htbp]
    \centering
    \includegraphics[width=\textwidth]{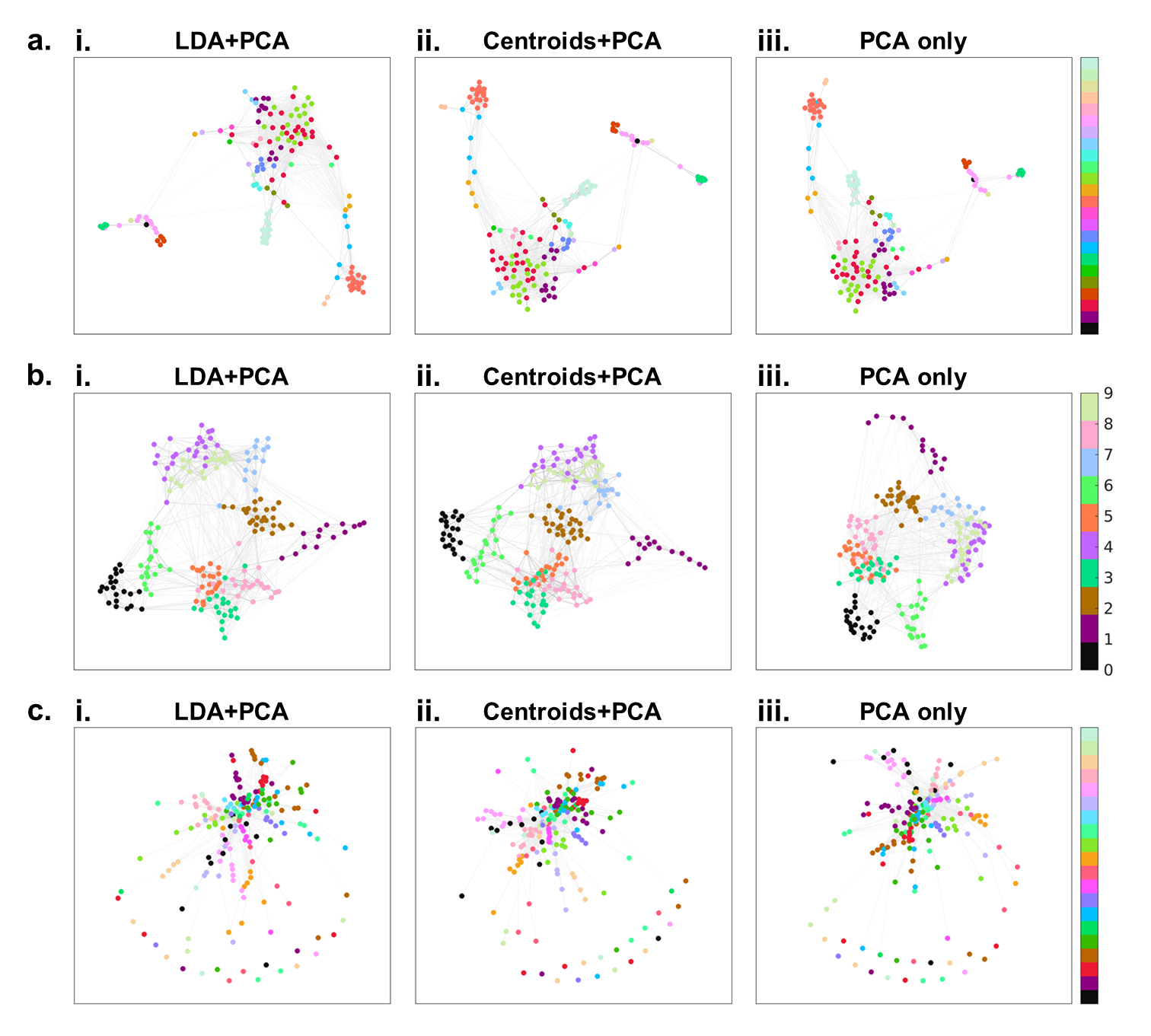}
    \vspace{-0.4cm}
    \caption[Comparison of force-directed layouts of ASTRICS similarity graphs generated using different local DR methods.]{\textbf{Comparison of force-directed layouts of ASTRICS similarity graphs generated using different local DR methods.} \textbf{(a)} Force-directed layouts of ASTRICS similarity graphs generated using the LDA+PCA (i), Centroids+PCA (ii), or PCA-only (iii) local DR algorithm for the Samusik01 CyTOF data seed clusters presented in \protect\hlcy{Figure}~\ref{fig:ASTRICS_Samusik01CyTOF}. \textbf{(b)} Force-directed layouts of ASTRICS similarity graphs generated using the LDA+PCA (i), Centroids+PCA (ii), or PCA-only (iii) local DR algorithm for the seed clusters of MNIST images of digits presented in \protect\hlcy{Figure}~\ref{fig:ASTRICS_MNIST}. \textbf{(c)} Force-directed layouts of ASTRICS similarity graphs generated using the LDA+PCA (i), Centroids+PCA (ii), or PCA-only (iii) local DR algorithm for the seed clusters of vectorized text documents from the \textit{20 Newsgroups} test dataset presented in \protect\hlcy{Figure}~\ref{fig:ASTRICS_20Newsgroups}a. For the newsgroups text document data, ASTRICS was performed in a 100-dimensional feature space obtained from global feature extraction by SVD. In all figure panels, nodes are coloured according to their ground-truth cluster assignments determined by majority voting of the data points within the corresponding seed clusters. For panels (a) and (c), refer to \protect\hlcy{Figures}~\ref{fig:ASTRICS_Samusik01CyTOF} and~\ref{fig:ASTRICS_20Newsgroups} respectively for the ground-truth cluster labels.}
    \label{fig:ASTRICS_comparison_of_DR_methods}
\end{figure}

So far, results have only been presented for the LDA+PCA algorithm for local DR in ASTRICS. To illustrate that the Centroids+PCA algorithm and even PCA-only can be valid alternatives to LDA+PCA for the local DR step of ASTRICS, \hlcy{Figure}~\ref{fig:ASTRICS_comparison_of_DR_methods} presents visualizations resulting from the use of each of the three local DR methods in a $K$-means--ASTRICS--force visualization pipeline. \hlcy{Figure}~\ref{fig:ASTRICS_comparison_of_DR_methods}a shows a force-directed layout of the ASTRICS similarity graph generated using (i) LDA+PCA, (ii) Centroids+PCA, or (iii) PCA-only as the local DR method in ASTRICS for the Samusik01 CyTOF data. \hlcy{Figures}~\ref{fig:ASTRICS_comparison_of_DR_methods}b and \ref{fig:ASTRICS_comparison_of_DR_methods}c make the same comparison of visualizations for the MNIST images of handwritten digits and for the test set of the \textit{20 Newsgroups} text documents respectively. For all three datasets, the three visualizations are qualitatively very similar (possibly with some rotations and/or reflections, which can be caused by different random initializations of the force-directed layout computation). 

Despite the comparable results of using the three local DR methods in ASTRICS in my pipeline for the three datasets used in this work, I would still favour LDA+PCA or Centroids+PCA over PCA-only from a theoretical standpoint. I believe that the use of PCA to initially reduce the dimensionality for a pair of clusters to prevent overfitting is a major reason for the similarity between the results of the three local DR methods. Even so, the PCA-only method never explicitly takes into account any separation between seed clusters, which means that it could, in theory, lead to overestimation of the similarity between seed clusters that are actually well resolved in a low-variance dimension. On the other hand, LDA+PCA and Centroids+PCA account for separation of clusters after the initial DR by PCA. Thus, LDA+PCA and Centroids+PCA should always theoretically yield results that are at least as good as PCA-only, although they retain more risk of overfitting. In addition, as noted earlier, LDA+PCA and Centroids+PCA can actually reduce the overall computation time required by ASTRICS compared to PCA-only, even though PCA-only certainly has the lowest computational complexity of the three local DR methods. Nonetheless, the results shown in \hlcy{Figure}~\ref{fig:ASTRICS_comparison_of_DR_methods} show that all three of the local DR methods that I propose are valid and the specific choice is not of critical importance.


\section{Discussion}

In this work, I have introduced a novel measure of similarity between clusters of HD data points and presented a fully automated method for its computation. I call this method \textit{Alpha Shape TRIangulation in loCal Subspaces} (ASTRICS). ASTRICS consists of two fundamental steps. The first is local DR using one of three methods that I have described and discussed: LDA+PCA, Centroids+PCA, or PCA-only, of which I favour LDA+PCA. The second step is computation of inter-cluster similarity based on triangulation of alpha shapes in locally determined 2- and 3-dimensional subspaces. I have primarily positioned ASTRICS as a tool that can be utilized as an intermediate step in a multistage pipeline for visualizing and clustering HD data. I have described such a pipeline (a seed--ASTRICS--clu/vis pipeline) in which ASTRICS generates a weighted graph from the clusters output by a clustering algorithm such as $K$-means and then force-directed layout and community detection are used to visualize and further cluster the nodes of the graph. Applying this pipeline to 39-dimensional CyTOF data, 400-dimensional images from the MNIST dataset, and newsgroups text document data with thousands of dimensions, I have demonstrated that ASTRICS is a suitable measure of inter-cluster similarity for a broad variety of HD data. Used as described in a seed--ASTRICS--clu/vis pipeline, ASTRICS facilitates analysis of HD data by enabling, in parallel, clustering and coarse-grained visualization that circumvent the ``curse of dimensionality''.

In spite of its demonstrated success, for example in improving clustering results for CyTOF data, I expect that my seed--ASTRICS--clu/vis pipeline could be improved by replacing the seed and final clustering steps with better or more specialist algorithms for a given clustering task. I chose $K$-means and the Louvain method for these two steps respectively to demonstrate the potential utility of ASTRICS because they are very simple, fast, popular algorithms. However, I note, for example, that very rare populations of interest in CyTOF data could be lost during seed clustering with $K$-means, which tends to return clusters of similar `spatial' size (in the sense of HD space). To prevent the loss of rare populations during $K$-means seed clustering, $K$ must be chosen large enough that (a) the returned clusters are not spatially larger than any of the rare populations and (b) there is a good probability that at least one data point in each rare population is selected during the initialization of cluster centroids. Alternatively, it might be preferable to replace $K$-means for the seed clustering by an algorithm tailored more towards identifying heterogeneous clusters, thus allowing smaller values of $K$ to be used for seed clustering. For example, a method such as PhenoGraph \cite{Levine2015phenograph}, which does not make any assumptions about the shapes or sizes of clusters, could be used to determine the seed clusters. The trade-off for using more advanced methods for the seed clustering will generally be increased computational time, however.

Similarly, for the final clustering step, the Louvain method could be substituted. The Louvain method is subject to a resolution limit \cite{Fortunato2007reslimit} that practically restricts the size and number of clusters that can be detected in a dataset. It also does not permit the final number of clusters to be specified, but in some cases this could be desirable. Multi-level Infomap \cite{Rosvall2011infomapM} is an alternative hierarchical community detection algorithm that is not subject to a resolution limit. Spectral clustering is another possibility, which includes a heuristic for selecting the best number of clusters. Nevertheless, as I have shown, the Louvain method still performed well. The performance of my pipeline even when pairing ASTRICS with very simple and efficient clustering algorithms provides a very strong baseline for the potential benefits of using ASTRICS. Even as the development of new clustering algorithms continues, ASTRICS can be paired with them to further enhance clustering.

Although I have demonstrated application of ASTRICS within a pipeline for visualizing and clustering a single set of HD data, I point out that ASTRICS has other utilities. Firstly, there is no requirement that the seed clusters all belong to the same dataset; the seed clusters could originate from multiple datasets as long as they share a common set of dimensions (i.e.~features) from which ASTRICS can proceed. Hence, ASTRICS followed by a similarity- or graph-based clustering algorithm could be used to aggregate existing clusters from multiple sources. Along similar lines, ASTRICS in isolation could be used simply as a method to score the similarities between clusters from one or multiple datasets. For example, if clinical biomedical data from multiple patients or from multiple time points of a longitudinal study of a single patient has already been clustered for each patient or time point individually, ASTRICS could be used to identify the most similar clusters between different patients or time points. Furthermore, ASTRI, the inter-cluster similarity computation component of ASTRICS without the local DR step, is a novel similarity measure in its own right. ASTRI can serve as an alternative to measures such as the discrete Bhattacharyya coefficient for quantifying the similarity between spatial point clouds. The development of ASTRICS was motivated by my own research project (which will be described in a separate paper) involving CyTOF and demanding improved methods for analyzing HD CyTOF data, but ASTRI could equally be applied to clusters from 2D or 3D spatial data. Indeed, the general seed--ASTRICS--clu pipeline that I have described here, but without the local DR step of ASTRICS (so a seed--ASTRI--clu pipeline), could be used to cluster spatial data.

Finally, I discuss the time complexity of ASTRICS. For a pair of clusters, this depends on computations in each of the following steps: preliminary outlier detection (preliminary step One), local DR (the CS step), and similarity computation using computational geometry (the ASTRI step). The time complexities of both the preliminary outlier detection and local DR steps are dominated by PCA or SVD. The outlier detection step potentially also requires nearest neighbour searches, but these have a lesser time complexity. Computation of alpha shapes determines the time complexity of the ASTRI step. For $n$ data points with $d$ dimensions, PCA or SVD can be implemented with a time complexity of $\bigO(\min\{n^2d,\,nd^2\})$. Alpha shapes can be computed in $\bigO(n\log n)$ time. Hence, the overall time complexity for ASTRICS to compute the similarity of two clusters is $\bigO(nt)$, where $n$ is the number of data points in the cluster pair and $t=\min\{nd,\,d^2\} + \log n$. 

However, the actual time complexity for applying ASTRICS to a set of $K>2$ clusters, as in a seed--ASTRICS--clu/vis pipeline, is difficult to ascertain because it depends on the specific partitioning of the data. Preliminary step Two, which has a time complexity of $\bigO(\min\{K^2d,\,Kd^2\})$ dominated by application of PCA to the $K$ cluster centroids, restricts the number of pairs of clusters to which ASTRICS must be applied. In the worst-case scenario the core ASTRICS algorithm would be performed $\bigO(K^2)$ times. The number of data points in each cluster depends on $K$ and the structure of the data, but on average each cluster would contain $N/K$ data points, where $N$ is the total number of points in the dataset. Therefore, we can estimate the time complexity of ASTRICS for an average pair of clusters by substituting $n=2N/K$ above. 
Multiplying the resulting average estimate by the worst-case number of cluster pairs and adding the complexity of preliminary step Two yields an estimated worst-case overall time complexity of $\bigO(KNt+Kds)$ where now $t=\min\{2Nd/K,\,d^2\} + \log N - \log K + \log 2$ and $s=\min\{K,\,d\}$. Over the range of suitable values for $K$, this is an increasing function of $K$. The computation times of the seed and final clustering steps of a complete seed--ASTRICS--clu pipeline also increase with $K$. As a result, there is a trade-off between speed and accuracy of my two-tiered pipeline built around ASTRICS for clustering and visualization. Larger values of $K$ will yield higher-purity seed clusters in the first tier and consequently better quality visualization and clustering results in the second tier, as long as the seed clusters are large enough for the local DR and computational geometry methods to be reliable and to avoid overfitting, but this comes at the cost of greater computation time. I emphasize, however, that ASTRICS is easy to parallelize for $K>2$ clusters because computations for different cluster pairs are independent. Therefore, the potentially high computational cost of ASTRICS can be mitigated by parallel computing.

A function for implementing ASTRICS in MATLAB (MathWorks) with parallel computing is freely available from \url{https://bitbucket.org/jscurll/astrics/src/master/}. As a fully automated method for computing inter-cluster similarities, it is free from subjective user bias. This includes freedom from subjective discretization of the data space, which would be required for existing statistical measures of similarity between two finite sets of points. Used in the ways that I have discussed, I believe that ASTRICS can become an effective tool to facilitate analyses of HD data, especially in the biomedical sciences.


\section*{Acknowledgements}
I thank Prof.~Daniel Coombs (Department of Mathematics, University of British Columbia) and Prof.~Michael R.~Gold (Department of Microbiology \& Immunology, University of British Columbia), my PhD supervisors, for supporting this work, and I acknowledge funding from the Canadian Cancer Society Research Institute (Innovation Grant 704254 to Daniel Coombs and Michael R.~Gold).


\section*{Competing interests statement}
The author declares no competing interests.


\newpage

\singlespacing

\bibliography{main}

\end{document}